\documentclass[useAMS,usenatbib]{mnras}

\usepackage{graphics}
\usepackage{amssymb}
\usepackage{amsfonts}
\usepackage{wasysym}
\usepackage{epsfig}
\usepackage{breakurl}

\usepackage{longtable}
\usepackage{times}
\usepackage{mathptmx}
\usepackage[usenames,dvipsnames]{color}
\usepackage{url}

\newcommand{\s}{\scriptsize}

\newcommand{\XMM}{{\em XMM-Newton}}

\newcommand{\roa}{{$\rho$\,Oph\,A}}
\newcommand{\rob}{{$\rho$\,Oph\,B}}

\newcommand\bz{$\langle B_{\mathrm z}\rangle$}
\newcommand\nz{$\langle N_{\mathrm z}\rangle$}
\newcommand\sbz{$\sigma_{\langle B_{\mathrm z} \rangle}$}

\title[The auroras of \roa]
{Evidence for radio and X-ray auroral emissions from the magnetic B-type star \roa}
\author[P. Leto et al.]
{P. Leto,$^{1}$ \thanks{E-mail: paolo.leto@inaf.it}
C. Trigilio,$^{1}$
F. Leone,$^{2}$
I. Pillitteri,$^{3}$
C. S. Buemi,$^{1}$
L. Fossati,$^{4}$
F. Cavallaro,$^{1}$
\newauthor L. M. Oskinova,$^{5,6}$
R. Ignace,$^{7}$
J. Krti\v{c}ka,$^{8}$
G. Umana,$^{1}$
G. Catanzaro,$^{1}$
A. Ingallinera,$^{1}$
\newauthor F. Bufano,$^{1}$
C. Agliozzo,$^{9}$
N. M. Phillips,$^{9}$
L. Cerrigone,$^{10}$
S. Riggi,$^{1}$
S. Loru,$^{1}$
\newauthor M. Munari,$^{1}$ 
M. Gangi,$^{1}$ 
M. Giarrusso$^{11}$
and J. Robrade$^{12}$
\\
$^{1}$INAF - Osservatorio Astrofisico di Catania, Via S. Sofia 78, 95123 Catania, Italy\\
$^2$Universit\'{a} di Catania, Dipartimento di Fisica e Astronomia, Sezione Astrofisica, Via S. Sofia 78, I-95123 Catania, Italy\\
$^{3}$INAF - Osservatorio Astronomico di Palermo, Piazza del Parlamento 1, 90134 Palermo, Italy\\
$^4$Space Research Institute, Austrian Academy of Sciences, Schmiedlstrasse 6, A-8042 Graz, Austria\\
$^5$Institute for Physics and Astronomy, University Potsdam, 14476 Potsdam, Germany\\
$^6$Kazan Federal University, Kremlevskaya Str 18, Kazan, Russia\\
$^7$Department of Physics \& Astronomy, East Tennessee State University, Johnson City, TN 37614, USA\\
$^8$Department of Theoretical Physics and Astrophysics, Masaryk University, Kotl\'{a}\v{r}sk\'{a} 2, CZ-611 37 Brno, Czech Republic\\
$^{9}$European Southern Observatory, Karl-Schwarzschild-Strasse 2, Garching bei M\"{u}nchen, 85748, Germany\\
$^{10}$Joint ALMA Observatory, Alonso de C\'{o}rdova 3107, Vitacura, Santiago, Chile\\
$^{11}$INFN, Laboratori Nazionali del Sud, Via S. Sofia 62, I-95123 Catania, Italy\\
$^{12}$Hamburger Sternwarte, University of Hamburg, Gojenbergsweg 112, D-21029 Hamburg, Germany
}
\begin{document}

\date{}

\pagerange{\pageref{firstpage}--\pageref{lastpage}} \pubyear{}

\maketitle

\label{firstpage}

\begin{abstract}
We present new ATCA multi-wavelength radio measurements (range 2.1--21.2 GHz) 
of the early-type magnetic star \roa, performed in March 2019 during 3 different observing sessions.
These new ATCA observations evidence a clear rotational modulation of the stellar radio emission 
and the detection of coherent auroral radio emission from \roa\ at 2.1 GHz.
We collected high-resolution optical spectra of \roa\ acquired by several instruments over a time span of about ten years.
We also report new magnetic field measurements of \roa\ that, together with the radio light curves
and the temporal variation of the equivalent width of the He\,{\sc i} line ($\lambda=5015$~\AA), were used to
constrain the rotation period and
the stellar magnetic field geometry. 
The above results have been used to model the stellar radio emission,
{ modelling that allowed us to constrain the physical condition of \roa's} magnetosphere.
Past \XMM\ measurements showed periodic X-ray pulses from \roa.  
We correlate the X-ray light curve with the magnetic field geometry of \roa.
The already published \XMM\ data have been re-analyzed showing that the X-ray spectra of \roa\
are compatible with the presence of a non-thermal X-ray component.
We discuss a scenario where the emission phenomena occurring at the
extremes of the electromagnetic spectrum, radio and X-ray,
are directly induced by the same plasma process.
We interpret the observed X-ray and radio features of \roa\ as having an auroral origin.
\end{abstract}


%
%
%
\section{Introduction}
Since \citet{babcock49} and \citet{stibbs50}, spectroscopic, photometric, and magnetic variabilities with the same period
that characterize chemically peculiar (CP) stars are understood in the framework of the 
Oblique Rotator Model (ORM). 
The ORM explains photospheric variability with the stellar rotation period as a consequence of a frozen, 
mostly dipolar, magnetic field (order of kG) not aligned with the stellar rotational axis.

The capability of stellar magnetic fields to trap out-flowing mass 
can explain the periodic variability in the wings of H${\alpha}$ 
\citep{Walborn1974,Leone1993a} and
the profiles of UV lines \citep{Shore1981}.
Near infrared photometry evidences accumulation of matter locked by the stellar magnetic field \citep{Groote1982}.
The rotational modulation of the C\,{\sc iv} and Si\,{\sc iv} UV lines observed on magnetic stars
gave evidence of  co-rotating hot gas \citep{Shore1988,Shore1990}.

The interaction of a radiatively driven wind with the stellar magnetic field has been widely explored
\citep{babel_montmerle97,ignace_etal98,cassinelli_etal02}.
In the presence of a large scale stellar magnetic field the radiatively driven wind is magnetically channeled.
Within the Magnetically Confined Wind Shock (MCWS) model framework \citep{babel_montmerle97},
the wind plasma arising from the hemispheres of opposite magnetic polarity
collides and shocks at the magnetic equator,
helping to explain the detectable thermal X-ray emissions from this class of stars.
The MCWS scenario
has been extensively analyzed and updated by MHD simulations
\citep{ud-doula_owocki02,ud-doula_etal06,ud-doula_etal08}.
The X-ray emission properties from a large sample of hot magnetic stars 
largely agree with this scenario
\citep{oskinova_etal11,naze_etal14,robrade16},
although some deviations were observed.

The dynamic interaction between the magnetically channelled
{ wind and stellar rotation plays a key role in} the accumulation
of plasma within the magnetospheres of such stars and, consequently,
on their typical observing features.  In cases of { fast rotators} stars
with extremely strong magnetic fields, the rotation balances the
gravitational infall of the magnetospheric plasma
\citep{maheswaran_cassinelli09}, leading to the formation of a large
centrifugally supported magnetosphere, versus the dynamical
magnetosphere extending up to the Kepler co-rotation radius.

Following the simplified hypothesis of completely rigid magnetic field lines, 
\citet{townsend_owocki05} developed the Rigidly Rotating Magnetospheric (RRM) model, where  
the presence of circumstellar plasma,  
forced to rigidly co-rotate with the star by the magnetic field,
can explain the rotational behavior of the H$\alpha$
emission observed in some fast rotating and strong magnetic early-type stars 
\citep{groote_hunger97, 
shultz_etal19}. 
In particular, the study of the typical H$\alpha$ signature of a RRM
has been used to classify
early type magnetic stars with centrifugal magnetospheres  \citep{petit_etal13,shultz_etal19b}.

In cases of extremely strong magnetic confinement and fast rotation,
the Rigid-Field Hydrodynamics (RFHD) simulations of the RRM model \citep{townsend_etal07}
predict gas heating 
{ at temperatures high enough to also produce} hard X-rays
at the edges of the centrifugal magnetosphere.
An update of the original MWCS model accounting for the MHD results
is given by the X-ray Analytic Dynamical Magnetosphere (XADM) model developed by \citet{ud-doula_etal14}.

\begin{figure*}
\resizebox{\hsize}{!}{\includegraphics{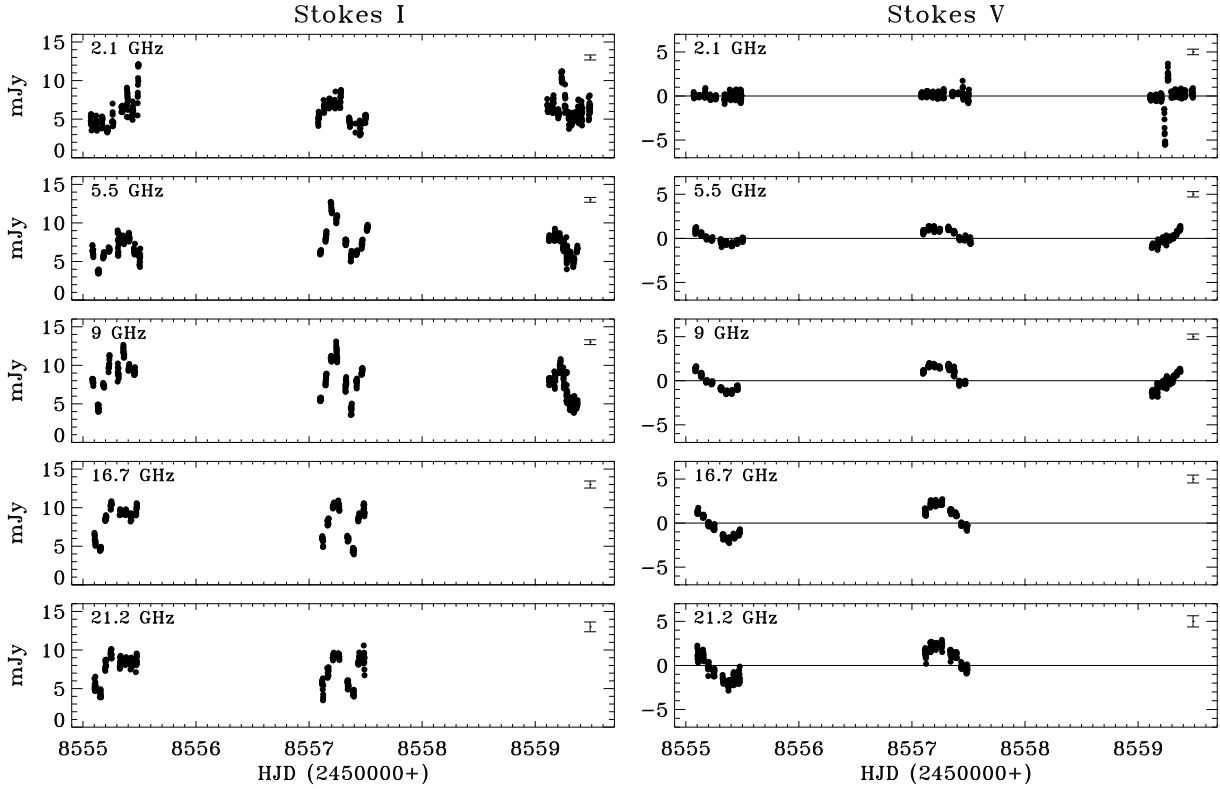}}
\caption{Multi-wavelength ATCA radio measurements of \roa.
Left panels refer to the  total intensity (Stokes\,{\em I}) measurements.
Right panels show the circularly polarized intensity (Stokes\,{\em V}).
The  representative error bars of the radio measurements are displayed in the right upper corner of each panel.} 
\label{fig_data}
\end{figure*}

A strong magnetic field and a plasma wind are key ingredients for establishing significant radio emission 
among early-type magnetic stars. 
The discovery of radio emission from these magnetic stars was made by \cite{Cassinelli1985}. 
Most of the known radio sources were
identified by \citet{drake_etal87,linsky_etal92}, and \citet{leone_etal94}. 
\cite{Leone1991} and \cite{Leone1993b} found that the radio emission of early type stars,
as well as the photometric, spectroscopic and longitudinal magnetic field strength variations,
are periodically variable with the stellar rotational period.

Following the framework outlined by the MCWS model,
the interaction of the stellar magnetic field with its radiatively driven stellar wind produces
gyro-synchrotron radio continuum emission,
as proposed by \citet{andre_etal88}.
The measurable effects induced by the stellar rotation
on the radio emission from a typical  
magnetosphere surrounding a hot magnetic star,
shaped like an oblique  
rigidly rotating magnetic dipole,
have been modeled  by \citet{trigilio_etal04} and \citet{leto_etal06}.

Some early-type magnetic stars are known sources of coherent radio emission 
\citep{trigilio_etal00, chandra_etal15,lenc_etal18,das_etal18,leto_etal19,das_etal19a,das_etal19b}.
Such coherent emission is
similar to the planetary Auroral Radio Emission (ARE) \citep{zarka98},
that arises from the magnetospheric regions  above the polar auroral ovals 
of the magnetized planets of the solar system, and are observed at infrared (IR), visible, ultraviolet (UV), 
and X-ray wavelengths \citep{badman_etal15}.

Stellar ARE from early-type magnetic stars 
was detected as wide band strongly circularly polarized radio pulses
arising from auroral rings above the magnetic poles \citep{trigilio_etal11}.
While the auroral signature in the UV spectrum of the prototypical
star CU\,Vir was not found \citep{krticka_etal19},
the search in X-rays gives promising results \citep{robrade_etal18}.

In this paper we present new multi-wavelength radio measurements of 
\roa\ (HD\,147933), a magnetic \citep{pillitteri_etal18} B2V~type star 
that has shown X-ray pulses with { an apparent period} of $\approx 1.2$ days \citep{pillitteri_etal17}.
\roa\ is a member of a binary system composed of two early B stars
separated by $\approx 3\arcsec$ and with
an extremely long orbital period (thousands of years; \citealp{malkov_etal12}).

\begin{table} 
\caption{ATCA observing log. Array configuration: 6A, Code: C3255.} 
\label{atca_log} 
\begin{center} 
\begin{tabular}{lcccc} 
\hline 
Date        & $\nu$            &Bandwidth        &Flux cal   &Phase cal       \\ 
                   & (GHz)                   & (GHz)       &                   &                         \\
\hline 
2019-Mar-12                    &2.1/5.5/9     &2                               &1934$-$638        &1622$-$253  \\ 
\smallskip
                                            &16.7/21.2   &2                              &1934$-$638        &1622$-$297  \\ 

2019-Mar-14                 &2.1/5.5/9     &2                                &1934$-$638        &1622$-$253  \\ 
\smallskip
                                        &16.7/21.2   &2                               &1934$-$638        &1622$-$297  \\ 

2019-Mar-16               &2.1/5.5/9      &2                              &1934$-$638        & 1622$-$253  \\ 
\hline             
\end{tabular}     
\end{center} 

\end{table} 

At the sky position of \roa,
the 1.4 GHz NRAO/VLA Sky Survey (NVSS, \citealp{condon_etal98}) detected a radio source of 10.8 mJy. 
Unfortunately the low spatial resolution of the NVSS ($45\arcsec$) was unable to resolve the double system,
but unpublished VLA radio { observations} of the \roa\ field (CODE: AK460),
performed at 1.4 GHz in array A configuration, 
reached a sufficient spatial resolution to resolve the stellar system, revealing that only \roa\ is a radio source.

In this paper we report new secure radio detections of \roa.
We also present new magnetic field measurements and high resolution optical spectra.
The collected radio and optical data sets (new and archival)
were used to infer the stellar rotation period
and to characterize the ORM of \roa.
The comparison of the observational features of \roa\ (radio and magnetic field)
with the already published X-ray measurements is discussed within the framework of the auroral emission model.

The structure of the paper is as follows: section \ref{obs} describes the observations and data analysis;
section \ref{sec:ephe} describes the updated ephemeris and rotation of \roa\ { and} section \ref{sec_radio_curve}
describes the radio light curves of \roa\ in different bands and its coherent and incoherent radio emission.
In sections \ref{beff_curve} and \ref{sec_magsph} we discuss the ORM geometry and 
the simulations of the multi-wavelength radio light curves, 
which { allowed us to} constrain the physical conditions of the magnetosphere of \roa.
In section \ref{sec_x_ray} the X-ray properties of \roa\ are discussed in relationship
with the modeling results of its radio emission.
In  section \ref{sec_are} the auroral radio emission is studied in accordance with the X-ray properties of \roa.
In sections \ref{discussion} and \ref{conclusion} we discuss our results and present our conclusions.

\section{Observations and data reduction}
\label{obs}
\subsection{Radio measurements}
\label{sec_atca}
In this paper we present new radio observations of \roa\
performed with the Australian Telescope Compact Array (ATCA)\footnote{The Australia Telescope Compact Array 
is part of the Australia Telescope National Facility which is funded by the Australian Government for
operation as a National Facility managed by CSIRO}.
The ATCA interferometer was used with its maximum allowed baseline length (6 km).
The observations were performed using the new broadband backend (CABB) system 
(bandwidth 2 GHz for each band). The selected bands were the L, C, X, U, and K bands, 
centered at: 2.1, 5.5, 9, 16.7, and 21.2 GHz.
For all the selected observing bands, 
the flux and bandpass calibrations were performed observing the standard calibrator PKS 1934$-$638.
The phase calibrator closest to \roa\ is PKS 1622$-$253. This source was used at the L, C, and X bands,
but it was too faint at the higher frequencies.
Thus, the adopted phase calibrator at the U and K bands was 
the next-closest calibrator: PKS 1622$-$297.

The target was observed in three different observing epochs during March 2019. 
During the third observing session, observations  were performed only at L, C and X bands, due to 
bad weather conditions that compromised the quality of the high-frequency observations.
The observing log is reported in Table~\ref{atca_log}.

The ATCA measurements  were edited and calibrated employing the {\sc miriad} software package.
Sub-bands contaminated by strong RFI were flagged (task {\sc blflag}).
During each observing run, \roa\ was observed by ATCA for $\approx 10$ hrs per band.
The new ATCA observations cover a large range of hour angles that allows us to obtain high-quality 
images at the sky position of \roa\ (tasks {\sc invert}, {\sc clean}, and {\sc restore}).
The measured map noise level is close to the nominal value for ATCA $10$-hrs-long observations,
namely: 0.005 mJy/beam at L band; 0.004 mJy/beam at C and X bands; 
0.008 mJy/beam at U band;  0.015 mJy/beam at K band.
The ATCA spatial resolution at $\nu \geq 9$ GHz is better then $3 \arcsec$ ($\approx 0.65 \arcsec$ at the highest observing frequency),
allowing us to confirm that \rob\ is undetected above the map noise threshold.
The non-detection of \rob\ ensures that the radio emission of \roa\ is not contaminated. 

The ATCA interferometer has a linear array design that precludes imaging using short-time observations. 
To measure the flux density time variation of \roa, both for the total intensity 
(Stokes\,{\em I}) and the circularly polarized intensity (Stokes\,{\em V}),
discrete Fourier transforms(DFT) of the complex visibilities at the source position were computed as function of time.
The {\em I} and {\em V} Stokes parameters were obtained combining
the right (RCP) and left (LCP) hand circularly polarized components of the electromagnetic wave
($I= ({\mathrm{RCP}}+{\mathrm{LCP}})/2$; $V=({\mathrm{RCP}}-{\mathrm{LCP}})/2$).
Such methods were used first by \citet{trigilio_etal08} to analyze the fully polarized pulses of CU\,Vir measured by ATCA and,
more recently, to analyze the dynamical radio spectrum of $\alpha$\,Cen, also with ATCA data \citep{trigilio_etal18}.

The adopted resolution time for the DFT procedure is 1 minute, with related uncertainties of: 
0.1 mJy/beam at L band; 0.09 mJy/beam at C and X bands; 
0.2 mJy/beam at U band;  0.4 mJy/beam at K band.
To take into account the flux density uncertainty of the adopted phase calibrators
($\approx 1 $ per cent for the selected observing bands),
one per cent of the \roa\ flux density (measured in each time bin) was added in quadrature to the above reported uncertainties.
The time-resolved ATCA radio measurements of \roa\ are displayed in Fig.~\ref{fig_data}
{ (the radio data are listed in Table~\ref{radio_data})}.
As one can see, the radio emission of \roa\ varies slowly over time,
except for the highly polarized radio transient
detected at 2.1 GHz during the last observing session,
which closely resembles the coherent pulses from CU\,Vir \citep{trigilio_etal00}.

\subsection{Magnetic field measurements}
\label{sec:mag_field_obs}
Following the first detection of a magnetic field for \roa\ \citep{pillitteri_etal18}, 
we collected additional magnetic field measurements 
primarily aiming at identifying 
the stellar magnetic field geometry. These observations have been
obtained using the 
FORS2\footnote{Based on observations collected at the European Southern Observatory under ESO programme ID 096.C-0159(A).} 
low-resolution spectropolarimeter \citep{appenzeller1998}, 
which is attached to the 
ESO/VLT~UT1 (Antu) of the Paranal Observatory (Chile). 
The data were taken in seven
epochs from July to September 2018 (see Table~\ref{tab:Bfield}), with a slit width of $0.4\arcsec$, 
using grism 600B. The choice of the grism and the slit width resulted in a resolving power of 
approximately 1700. The spectra cover the range 3250--6215\,\AA, which includes all Balmer
lines except H$\alpha$, and a number of He lines. For each epoch of observation, 
the star was observed with a sequence of spectra (see column three of Table~\ref{tab:Bfield}) 
obtained by rotating the quarter waveplate alternatively from $-$45$^\circ$ to $+$45$^\circ$
every second exposure (e.g., $-$45$^\circ$, $+$45$^\circ$,
$+$45$^\circ$, $-$45$^\circ$, $-$45$^\circ$, $+$45$^\circ$,
$+$45$^\circ$, $-$45$^\circ$). The exposure times and obtained
signal-to-noise ratios (S/N) per pixel calculated around 4950\,\AA\
of Stokes $I$ are listed in Table~\ref{tab:Bfield}.

\begin{table}
\caption[ ]{FORS2 and Narval observing log of the \bz\ measurements.}
\label{tab:Bfield}
\begin{footnotesize}
\begin{tabular}{@{}lc c c c@{}}
\hline
Date & $<\mathrm{HJD}> $                  & S/N                  &~~ \bz\ (G)                   &~~ \bz\ (G)           \\ 
         & ($2450000+$)                           &                        &~~{Hydrogen}        &~~{All}          \\ 
\hline
2014-Jul-10$^{\dag}$ &   { 6849.395\s{(0.02)~~}}  &  ~~900     &    ~~~-                           &    $-143${$\pm235$} \\ 
2017-Jul-17$^{\dag\dag}$                          &  { 7951.543\s{(0.004)}}   &  2400   &    $-283${$\pm107$}  &    $-128${$\pm68~~$}     \\ 
2017-Aug-11$^{\dag\dag}$                        &  { 7976.573\s{(0.003)}}   &  3100   &	~~~485{$\pm84~~$}      &      ~~~404{$\pm55~~$}    \\ 

2018-Jul-11                          & { 8310.60\s{(0.04)}~~~~~}  & 2400   &    ~~~340{$\pm102$}  &    ~~~390{$\pm65~~$}     \\ 
2018-Jul-31                          & {8330.676\s{(0.006)}}   & 2500   &	~~~~18{$\pm94~$}      &      ~~~~~~~6{$\pm61~~$}    \\ 
2018-Aug-13                        & {8343.533\s{(0.002)}}   &   ~~800    &    ~~~531{$\pm330$}  &     ~~~~~46{$\pm204$}  \\ 
2018-Sep-05                        & { 8366.533\s{(0.002)}}   & 1600   &   $-138${$\pm153$}  &   ~~$-56${$\pm97~~$}  \\ 
2018-Sep-06                        & { 8367.55\s{(0.01)}~~~~~}  &   ~~700    &    ~~~889{$\pm370$}  &    ~~~654{$\pm242$}\\ 
2018-Sep-10                        & {8371.598\s{(0.001)}}   &   ~~700    &   $-232${$\pm326$}  &  $-167${$\pm237$} \\ 
2018-Sep-19                        & { 8380.535\s{(0.007)}}   &  1300  &   $-410${$\pm215$}  &   $-320${$\pm144$} \\ 
\hline
\end{tabular}
\end{footnotesize}
\begin{list}{}{}
\item[]{The S/N per pixel of Stokes $I$ calculated at about 4950\,\AA\ over a wavelength range of 100\,\AA.}
\item[$^{\dag}$]{Measurement retrieved from the Narval archive.}
\item[$^{\dag\dag}$]{FORS2 measurements already published by \citet{pillitteri_etal18}.}
\end{list}
\end{table}


Similar to the data presented in \citet{pillitteri_etal18}, we reduced and analysed the FORS2 spectra 
employing the pipeline described in \citet{fossati2015}, which is based on the algorithms and 
recommendations given by \citet{bagnulo2012,bagnulo2013}. 
We then derived the surface averaged longitudinal magnetic field \bz\ and its uncertainty \sbz\ 
using the method first described in \citet{bagnulo2002} for Stokes $V$ spectra. 
We further calculated the diagnostic null profile $N$, and hence \nz, following the formalism of 
\citet{bagnulo2009}. 
We computed both \bz\ and \nz\ using either the hydrogen lines or the whole spectrum. 
Table~\ref{tab:Bfield} summarizes our results.

On average the 2018 measurements
were performed using spectra with lower quality
than those obtained in 2017 \citep{pillitteri_etal18},
resulting in \bz\ measurements affected by bigger errors.
The origin of the larger uncertainties can be explained as follows.
The 2017 observations  presented by \citet{pillitteri_etal18} were
conducted in visitor mode, implying that the exposure times
could be fine tuned for the sky conditions to achieve a S/N
as high as possible. Furthermore, 
in visitor mode the data could be obtained
using the E2V CCD detector, which is more sensitive in the blue
where most of the hydrogen Balmer lines are located, thus maximizing
the signal in the spectral region carrying information on the
stellar magnetic field. Since the purpose was to measure \bz\ across
a wide time range to follow its rotational modulation,
the 2018 observations were obtained 
in service mode. The exposure times could be less
ideally adjusted to the sky conditions. More importantly, these
observations had to be carried out using the MIT CCD detector, which
is more sensitive in the red, leading to a lower signal-to-noise
ratio across the region covered by the hydrogen Balmer lines.

High resolution spectra of \roa\ were also acquired by the NARVAL spectropolarimeter, 
mounted at the Bernard Lyot Telescope,
see PolarBase database\footnote{\url{http://polarbase.irap.omp.eu}.}
{ \citep{petit_etal14,donati_etal97}.}
These spectra were acquired almost continuously for about 1.5 hrs, for a total of 40 different exposures.
The single exposures were averaged over the whole observing time.
Using all the available lines, following the method described in \citet{fossati_etal15a}, 
we obtain the further \bz\ measurement listed in Table~\ref{tab:Bfield}.

\subsection{High resolution spectroscopy}
\label{spectra}
Spectroscopic observations of \roa\ were carried out with the Catania Astrophysical Observatory 
Spectropolarimeter (CAOS) which is a fiber fed, high-resolution, cross-dispersed echelle spectrograph 
\citep{Leone2016} installed at the Cassegrain focus of the 91~cm 
telescope of the âM. G. Fracastoroâ observing station of the Catania Astrophysical Observatory (Mount Etna, Italy).

The CAOS spectra were obtained between May to August 2018,
for a total of 8 observing epochs; 
their exposure times have been tuned in order to obtain
a S/N of at least 100\ in the continuum across the 3900--6800 {\AA} spectral range, 
with a resolution of R = 45000, as measured from ThAr and telluric lines.

The reduction of all spectra, which included the subtraction of the bias frame, trimming, correcting for the 
flat-field and the scattered light, extraction of orders, and wavelength calibration, was done using the 
{\sc noao/iraf}  packages {\sc ccdred} and {\sc specred}. 
Given the importance of Balmer lines in our analysis, we paid much more attention to the normalization 
of the corresponding spectral orders. 
In particular, we divided the spectral order containing Balmer lines by a pseudo-continuum obtained from combining 
the continua of the previous and subsequent echelle orders, as already outlined by \citet{catanzaro15}. 
The {\sc iraf} package {\sc rvcorrect} was used to determine the barycentric
velocity and correct the observed radial velocities for the Earth's motion.

Unpublished high resolution spectra of \roa\ 
were found in the archives from HiRes@KECK, UVES@UT1, HARPS@3.6ESO, ESPaDOnS,@CFHT,
NARVAL@Telescope Bernard Lyot,
and HARPS-N@TNG.
These spectra have been retrieved and reduced by using the method described above.

The close proximity of \roa\ to its stellar companion \rob, 
{ with a similar} spectral type, might be a critical issue. 
Observations performed when \roa\ was at low elevation above the horizon of the site 
and the possible worse seeing condition might affect reliability of the measurements.
Therefore,
to search for possible light contamination { within \roa's} spectra 
produced by the light of its companion,
we inspected the spectral { profiles} of the helium lines,
that are likely present within the spectra of both components.
We found absolutely symmetric profiles of the examined lines,
except for the spectra acquired by CAOS, ESPaDOnS, and NARVAL.
Given these concerns, we took exceptional care with the
use of these spectra.

\begin{figure}
\resizebox{\hsize}{!}{\includegraphics{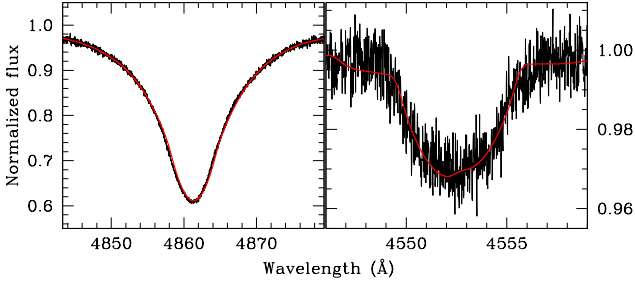}}
\caption
{The left panel shows the comparison between observed (HARPS-N) and computed H$\beta$. The right panel 
shows the comparison between observed and synthetic Si\,{\sc iii} line at $\lambda = 4552.622$~\AA.}
\label{high_res_spec}
\end{figure}

In B-type stars, Balmer line profiles are sensitive both to effective temperature and surface gravity. 
For the analysis of Balmer lines, we used the HARPS-N@TNG spectrum 
where there is no evidence of light contamination by \rob.
The approach we used in this paper is to minimize the difference between observed and synthetic Balmer lines profiles. 
Following \citet{catanzaro2016}, as goodness-of-fit test we used the ratio:
\begin{displaymath}
\chi^2 = \frac{1}{N} \sum \left(\frac{I_{\rm obs} - I_{\rm th}}{\delta I_{\rm obs}}\right)^2
\end{displaymath}
where $N$ is the total number of points, $I_{\rm obs}$ and $I_{\rm th}$ are the intensities of the 
observed and computed profiles, respectively, and $\delta I_{\rm obs}$ is the photon noise. 
Atmospheric models obtained with {\sc atlas9} \citep{kur93} use precomputed line opacities in the 
form of Opacity Distribution Functions (ODFs), that are tabulated for multiples of the solar metallicity and 
for various microturbulent velocities.
Atmospheric models have been computed by using the {\sc atlas9} code \citep{kur93} in 
Local Thermodynamic Equilibrium (LTE) approximation, while a stellar spectrum was 
synthesized using the spectrum synthesis code {\sc synthe} \citep{kur81}.

From our procedure we estimated 
$T_{\rm eff} =20800$ K and $\log g =4.0$. 
These values of effective temperature (reported in Table~\ref{par_star}) 
and surface gravity are in agreement, within the experimental errors, with 
the values given by \citet{pillitteri_etal18}. 
The synthetic profiles were { calculated} using almost solar abundance, 
as reported by \citet{pillitteri_etal18}. 
The comparison between the theoretical and observed H$\beta$ line profiles 
is shown in the left panel of Fig.~\ref{high_res_spec}.

To measure $v \sin i$, we matched synthetic line profiles from {\sc synthe} to a number of metal lines; the best 
fit was obtained with $v \sin i=200$ km~s$^{-1}$ (Table~\ref{par_star}),
which is still well in accordance with the value retrieved by \citet{pillitteri_etal18}. 
For our calculation we neglected the contributions of other velocity fields, as for
instance macroturbulence, since the line profiles are dominated by the stellar rotation. 
For the sake of clarity, we show in the
right panel of Fig.~\ref{high_res_spec} the result of the fitting of the Si\,{\sc iii} line fit at 
$\lambda=4552.622$~\AA, where only rotational broadening has been considered.

\begin{table}
\caption[ ]{\roa\ stellar parameters}
\label{par_star}
\footnotesize
\begin{tabular}[]{@{}lc@{}c@{}}
\hline
$P_{\mathrm {rot}}$ (days)   &$0.747326 \pm 0.000002$       & This work                                 \\
$d$ (pc)          & $140\pm4$                      & \citet{gaia_dr2}            \\
$T_\mathrm{eff}$ (K)          & $20800\pm500$                      & This work            \\
$M_{\ast}$ (M$_{\odot}$)  & $8.2^{+0.8}_{-0.7} $                            &  \citet{pillitteri_etal18}               \\
$\log g$ (cgs)                                & $4.0\pm0.1$                & \citet{pillitteri_etal18}            \\
$\log L/{\mathrm{L_{\odot}}}$               & $3.61 ^{+0.17} _{-0.16}$                & \citet{pillitteri_etal18}            \\
$u^{\dag}$                                        &0.3                   & \citet{claret_etal11}            \\
$R_{\ast}$ (R$_{\odot}$)   & $4.5 \pm 0.6  $    & \citet{pillitteri_etal18}            \\

$R_{\mathrm {eq}}$ (R$_{\odot}$)   & $5.2 \pm 0.7  $    & { This work}            \\

{$v \sin i$} (km s$^{-1}$)                & {$200 \pm 10 $}       &    This work                                \\
$W^{\dag\dag}$           & $0.64 \pm 0.09$              &    { This work}                                \\

\hline
\multicolumn{3}{l}{{ORM parameters}}\\
\hline
{$i$} (degree)                & $35 ^{+8} _{-6}$       &    { This work}                                \\
{$\beta$} (degree)                & $78 ^{+5} _{-8}$       &    { This work}                                \\
$B_{\mathrm p}$ (G)    & $2700 ^{+900} _{-700}$      &   { This work}                                \\

\hline
\multicolumn{3}{l}{{Magnetosphere parameters}}\\
\hline
$R_{\mathrm{A}}$ ($R_{\ast}$)   & 8--12                   &    This work                                \\
$v_{\infty}$ (km s$^{-1}$) & 1500      &    \citet{krticka14}                               \\
$\dot{M}$ (M$_\odot$\,yr$^{-1}$) & $2.6\times10^{-10}$--$2.1\times10^{-9}$~      &    { This work}                       \\
$\dot{M}_{\mathrm{act}}$ (M$_\odot$\,yr$^{-1}$) & ~$1\times10^{-11}$--$1.4\times10^{-10}$      &    { This work}                       \\
$R_{\mathrm{K}}$ ($R_{\mathrm{eq}}$)             & $1.35 \pm 0.1$              &    { This work}                                \\
\hline
\end{tabular}

\begin{list}{}{}
\item[{$\dag$}]{Limb darkening coefficient in the visual band.}
\item[{$\dag\dag$}]{Critical rotation parameter.}
\end{list}
\end{table}

\section{Ephemeris}
\label{sec:ephe}

\begin{figure}
\resizebox{\hsize}{!}{\includegraphics{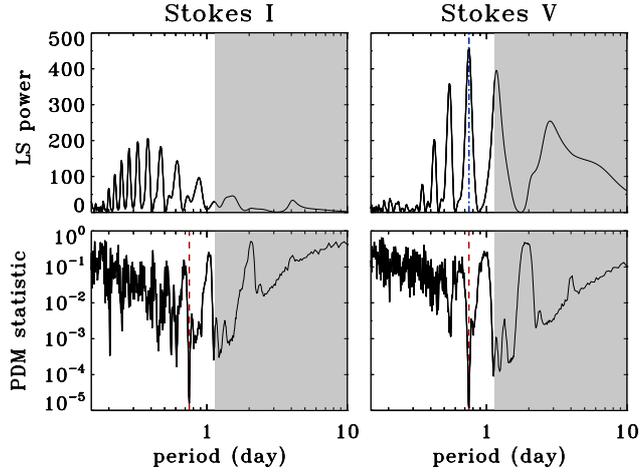}}
\caption{Period search of \roa\ based on the analysis of the Stokes\,$I$ (right panels) and Stokes\,{\em V} (left panels) radio measurements.
Top panels: Lomb-Scargle (LS) periodogram.
Bottom panels: phase dispersion minimization (PDM) method.
The vertical red dashed lines highlight the common period of $\approx 0.748$ found by the PDM method, both Stokes\,$I$ and $V$;
the  vertical blue dot-dashed line highlights the period of $\approx 0.749$ found by the LS method in the case of the Stokes\,$V$ only.
Grey regions refer to periods that are not compatible with the projected rotation velocity of \roa.}
\label{per_search}
\end{figure}

The search for
periodic variation of the radio emission of \roa\ was performed by using two distinct methods,
the Lomb-Scargle (LS) periodogram \citep{lomb76,scargle82} and
the phase dispersion minimization (PDM) method \citep{stellingwerf78}. 
A period estimated by the PDM method  will be significant when the PDM statistic is close to zero,
whereas for the LS method, a significant period is related to a maximum of the Lomb-Scargle periodogram.
The results of the period search applied to the radio measurements of \roa\ are shown in Fig.~\ref{per_search}.

The two methods applied to the Stokes\,$V$ data set (Fig.~\ref{per_search}
right panels) clearly indicate the existence of a period able to
phase fold the time series of the Stokes\,$V$ radio measurements.
The two separate methods found very similar periods, with PDM at $\approx 0.748$
days, and LS at $\approx 0.749$ days.  A period of  $\approx
0.748$ days was also found by the PDM method when applied to the
Stokes\,$I$ data set (bottom left panel of Fig.~\ref{per_search}),
whereas the LS method failed to identify a clear periodicity  (top left
panel of Fig.~\ref{per_search}). This is probably due to the 
periodicity of the Stokes\,$I$ data not being a simple sinusoid. In fact the
LS periodogram is an extension of the Fourier method applied to data that are 
non-uniformly spaced with time. On the other hand, the PDM statistic
(well-suited to search for non-sinusoidal variability) found the same period
from both the Stokes\,$I$ and $V$ radio measurements.

{
As is common in the case of magnetic stars,
the cyclic variability of \roa's radio emission
is a direct consequence of stellar rotation, }
hence the variability period coincides
with the stellar rotation period ($P_{\mathrm{rot}}$). 
The measured projected rotation velocity of \roa\ (Sec.~\ref{spectra}) put strong constraints on the allowed rotation periods.
Assuming a viewing inclination of $90^{\circ}$ to the spin axis,
the relation $v \sin i = 2\pi R_{\ast} \sin i / P_{\mathrm {rot}}$  indicates that 
the rotation period of \roa\ has to be lower than 1.135 days,
using $R_{\ast}=4.5$ R$_{\odot}$ \citep{pillitteri_etal18}.
The secondary peaks related to implausible periods fall in the grey regions of the periodograms of Fig.~\ref{per_search}. 
The time base of the radio measurements is $\approx 5$
days for the C and X bands and $\approx 3$ days for the U, K and L bands. 
The L-band measurements performed during the third observing
run were not used for the period search due to the detection 
{ of a highly} circularly polarized fast radio transient 
(see top right panel of Fig.~\ref{fig_data}).  
The uncertainty in the rotation period is about $\approx 0.006$ days ($8$ minutes).  
This uncertainty value was estimated from varying $P_{\mathrm {rot}}$ until
the $\chi^2$ of the sinusoidal fit of the { phased folded Stokes\,$V$ radio data increased} by one unit.

The radio measurements produce a rotation period that is significantly different with respect 
to the observed time separation (1.205 days) between the two X-ray pulses observed by \XMM\ in 2016. 
This is a clear indication that 
the X-ray emission is not simply related to the stellar rotation.

To better refine the precision of \roa's rotation period suggested by the modulation of its radio emission,
we searched for other observing features that are variable as a function of the stellar rotation.
The early-B type stars that are characterized by a well-ordered magnetic field topology
are commonly associated with an inhomogeneous surface distribution of the helium
\citep{kochukhov_etal11,oksala_etal15}.
We examined
the equivalent widths (EWs) of the He\,{\sc i} line at 
$\lambda=5015$~\AA~in the high resolution spectra acquired between 
the years 1997 and 2018 (Sec.~\ref{spectra}). 
The selected line is common to all the analyzed spectra and
we measured a significant variability of its EW (see Table~\ref{ew_data}).
When spectra acquired over long exposure time are available,
we obtain EW measurements from the spectrum averaged over exposure $\approx 20$ minutes long.
The measured effective magnetic field of \roa, listed in Table~\ref{tab:Bfield}, is also variable.
To determine the variability period, we applied the LS method to the radio (Stokes\,$V$) and optical data sets.
The observables retrieved by the optical data, 
that are assumed variable as a consequence of the stellar rotation and used to refine the period of \roa,
are the He\,{\sc i} line EW and \bz.
Since the origin of the X-ray emission is not fully clear,
we excluded the X-ray measurements { from the period analysis.
The behavior of \roa\ at the X-rays will be examined later}.
After the normalization of each periodogram { to the peak} value,  
we produced a new periodogram as the product of all single periodograms. 
The choice to normalize any single periodogram to its maximum was equivalent to assigning the same weight 
to the different data sets. 
\begin{figure}
\resizebox{\hsize}{!}{\includegraphics{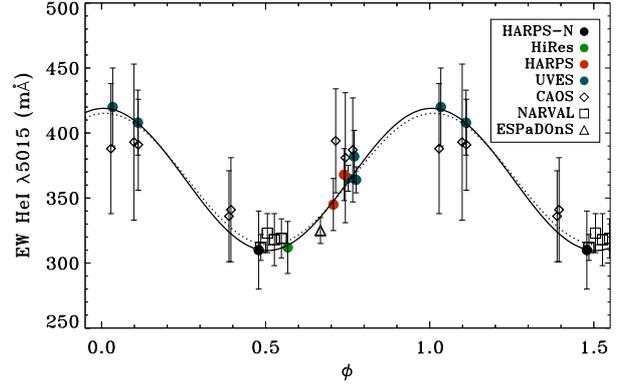}}
\caption{{ EW curve} of the He\,{\sc i} line at $\lambda=5015$ \AA.
The different symbols used to show the EW measurements refer to the listed instruments.  
The open symbols refer to \roa\ measurements suspected of light contamination 
by \rob, see text. Dotted line: sinusoidal fit of the whole data set.
Solid line: fit of the not contaminated measurements only.}
\label{fig_ew_hip}
\end{figure}

\begin{figure*}
\resizebox{\hsize}{!}{\includegraphics{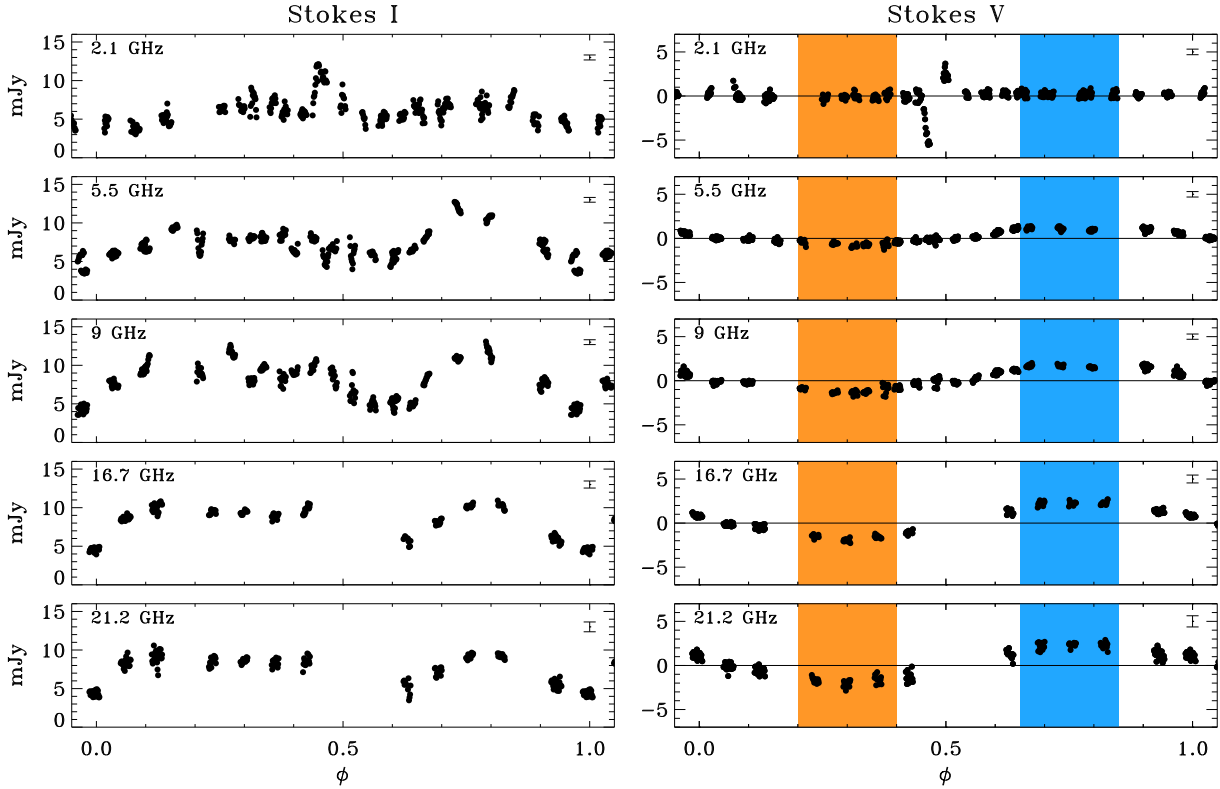}}
\caption{Rotational variability of the multiwavelength \roa\ radio emission.
Left panels:  total intensity (Stokes\,{\em I}).
Right panels: circularly polarized intensity (Stokes\,{\em V}).
The vertical orange bars ({ centred} 
at $\phi=0.3$) highlight the range of phases where Stokes\,{\em V} 
is clearly negative at $\nu \geq 5.5$ GHz;
the light-blue bars  ({ centred} at $\phi=0.75$) refer to the phase range with Stokes\,$V >0$.}
\label{l_fase}
\end{figure*}

As discussed in Sec.~\ref{spectra}, some high resolution spectra of \roa\ show evidence of light contamination from \rob.
To check for a systematic variability effect, we also performed the period search analysis 
excluding the EW measurements involving the contaminated spectra.
The exclusion had negligible effect on the period determination for \roa.
Fixing the zero-point of the stellar phases with the He\,{\sc i} EW maximum, 
we derive the following ephemeris:
\begin{equation}
\label{effemeridi}
{{\mathrm {HJD}}=2~450~672.36(1)+0.747326(2)E ~{\mathrm{(days),}}}
\end{equation}
where the error related to the last digit is given in brackets. 
The period uncertainty was estimated by 
measuring the difference between the period
where the maximum of the periodogram peak occurs
and the period related to the maximum of the gaussian function that better fit the periodogram peak.
The uncertainty related to the zero point of the ephemeris was retrieved by the errors of the sinusoidal fit parameters.
The rotational modulation of the EW measurements, phase folded using the above reported ephemeris, 
is shown in Fig.~\ref{fig_ew_hip}. 
It is evident  a clear sinusoidal variation
of the { un-contaminated data (measurements marked by filled symbols)},
variability that remains almost unchanged also using the suspected spectra (open symbols).
The apparent rotational variability of the He\,{\sc i} line is indicative of the presence of chemical spots on the surface of \roa,
similar to the case of HD\,37479 \citep{oksala_etal15}.

\section{The radio light curves of \roa}
\label{sec_radio_curve}

The multi-wavelength ATCA radio measurements of \roa\
were phase folded using the ephemeris reported in Eq.~\ref{effemeridi},
with radio light curves displayed in Fig.~\ref{l_fase}.
It is evident that the new ATCA observations
cover almost the full stellar rotation period of \roa\ (see Sec.~\ref{sec:ephe}).
The total intensity and the circularly polarized radio emission of \roa\ are modulated by the stellar rotation. 

The light curves for the Stokes\,{\em I} measurements evidence a similar shape at $\nu \geq 5.5$ GHz.
In particular, the measurements at $\nu=5.5$ and 9 GHz, that well sampled the whole stellar rotation period,
display two clear minima.
Further, the Stokes\,{\em V} emission shows that the sign of the circular polarization of the electromagnetic waves
changes as \roa\ rotates.
The corresponding variation amplitude grows as the observing frequency increases.

The measurements at the lowest radio frequency ($\nu = 2.1$ GHz) show a less clear rotational modulation,
both for Stokes\,{\em I} and {\em V}.
Instead an increase of the flux density confined within a narrow range of phases, 
close to $\phi \approx 0.5$, is evident.
The corresponding Stokes\,{\em V} measurements 
show that the circularly polarized
emission at $\nu =2.1$ GHz
is well detectable. Further, this event is characterized by the inversion of the polarization direction 
within the narrow range of covered rotational phase.

\subsection{Incoherent radio emission}
\label{radio_incoe}
The behavior of the radio emission from \roa\ is well in accordance with 
optically thick radio emission from  the stellar magnetosphere.
The detection of a clear rotational modulation and circularly polarized radio emission from \roa\
suggests that the non-thermal incoherent gyro-synchrotron emission mechanism is in operation.
A population of mildly relativistic electrons that move within the magnetosphere of \roa\
produce a continuum radio spectrum. This is the typical emission mechanism 
arising from hot magnetic stars surrounded by a stable co-rotating magnetosphere.

The incoherent multi-wavelength radio light curves of the well studied early-type magnetic stars,
CU\,Vir, 
HD\,37479, 
HR\,7355, 
and HR\,5907 \citep{leto_etal06,leto_etal12,leto_etal17a,leto_etal18},
were modeled by using a 3D model for the gyro-synchrotron emission
from a dipole shaped stellar magnetosphere \citep{trigilio_etal04}. 
In the framework of the MCWS model,
the magnetically confined wind leads to accumulation
of thermal matter within the magnetospheric regions where the magnetic field lines are closed (the ``inner magnetosphere").
Far from the star, near the Alfv\'{e}n radius ($R_{\mathrm {A}}$), the magnetic field no longer dominates the ionized trapped matter. 
In the resulting current sheets, electrons can be accelerated up to relativistic energies.
These non-thermal electrons, moving within the ``middle magnetosphere", radiate by the incoherent
gyro-synchrotron emission mechanism.
As a consequence of the ORM, the projected area of the radio source will be variable
as a function of rotation phase,
for both the total and the circularly polarized intensities.

It is commonly observed in early-type magnetic stars, 
that the non-thermal radio emission level 
enhances when the stellar magnetosphere shows mainly the polar regions 
(stellar orientations coinciding with the maxima of the effective magnetic field curve).
This is because the gyro-synchrotron emission mechanism is strongly sensitive 
to the magnetic field strength and orientation \citep{ramaty69,klein87}, 
which vary with the distance from the stellar surface.
Hence, radiation within a specific radio frequency band will be mainly emitted in a well
localized layer of the magnetosphere. 
Higher frequency emission is generated close to the
stellar surface, where the field strength is higher, while lower frequencies probe regions farther out.
The electromagnetic waves produced by
the gyro-synchrotron emission mechanism are also partially circularly polarized.
The circular polarization fraction and the corresponding polarization sign 
are a function of the average orientation of the magnetic field vectors (with respect to the line of sight)
within the magnetospheric regions where the radio emission of a fixed frequency predominantly originates.

At the stellar surface, the average magnetic field vector orientation is related to the \bz\ value.
The correlation between the effective magnetic field curve and 
the circular polarization fraction of the incoherent radio emission
has been clearly observed in many cases 
\citep{trigilio_etal04,bailey_etal12,leto_etal06,leto_etal12,leto_etal17a,leto_etal18,leto_etal19}.
Stellar orientations characterized by magnetic field lines mainly oriented toward the observer
(northern magnetic hemisphere dominant)
are related to gyro-synchrotron radio emission 
of mainly right-handed circular polarization (Stokes~$V$ positive).
Conversely, the radio emission is left-hand polarized (Stokes~$V$
negative) when the southern magnetic hemisphere is visible.  Moreover,
as a consequence of the radial dependence of the stellar magnetic
field, the height of the source region (where the radio emission
at a well fixed frequency mainly originates) affects the measured
fraction of the circularly polarized emission.  In fact, for a
simple dipolar field topology, radio emission is produced in a
region where the field lines are almost aligned.  Far from the star,
the magnetic field lines are curved, and regions with magnetic field
vectors of opposite polarities will contribute to the integrated
radio emission.  This causes depolarization, with consequent decrease
of the measured circular polarization fraction of the radio emission
arising far from the star (namely the lowest radio frequencies).

\begin{figure}
\resizebox{\hsize}{!}{\includegraphics{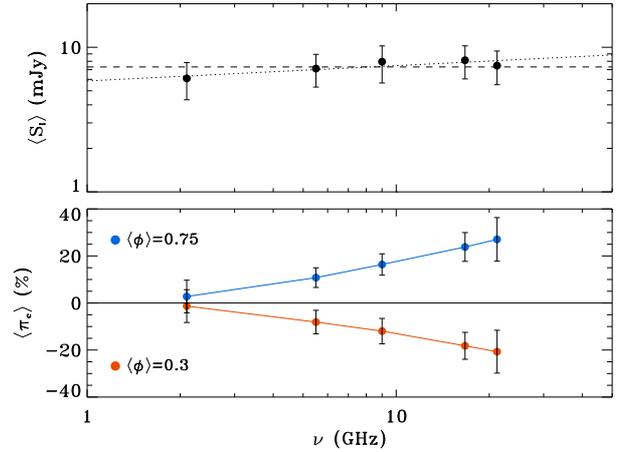}}
\caption{Top panel: average radio spectrum of \roa\,
fitted by using a power-law, (dotted line),
and a perfectly flat relationship (dashed line).
The error bar is the standard deviation of the rotationally modulated multi-wavelength radio measurements.
Bottom panel: spectrum of the fraction of the circularly polarized emission.
The two spectra were obtained separately by averaging the ATCA measurements performed
during the two ranges of rotational phases ($0.2$ large) showing Stokes\,{\em V} measurements always with the same sign.
}
\label{fig_spec}
\end{figure}

In the case of \roa\ there are wide ranges of rotational phases 
during which the Stokes\,{\em V} measurements have the same sign.  
The range of phases ($\Delta \phi \approx 0.2$)
mainly characterized by radio emission that is left-hand circularly
polarized (LCP) are highlighted in the right panels of Fig.~\ref{l_fase}
by the orange vertical bars; the blue bars refer to the phases
where the measured radio emission is RCP.
The spectral behavior of the circularly polarized emission has been
studied analyzing the average $\pi_{\mathrm c}$ 
(${\mathrm {Stokes}~V}/{\mathrm {Stokes}~I}$) calculated separately
within these phase ranges where the flux level and the polarization
direction remains almost constant.
The bottom panel of Fig.~\ref{fig_spec} shows the spectral dependence
of the fractional circularly polarized emission of \roa.  
The strength of $\pi_{\mathrm c}$ grows as the observing radio frequency increases.  
The highest fraction of the circularly polarized emission is measured at the K band 
($\nu=21.2$ GHz), ranging from $\approx -20$ per cent to $\approx +30$ per cent.
This is a clear indication that the higher frequencies mainly originate from magnetospheric layers
close to the stellar surface.
In accordance with a dipolar magnetic field topology,
the magnetic field vectors are almost radially oriented with respect to the stellar surface
in the layers that mainly radiate the higher radio frequencies.
This explains the frequency dependent effect of the strength of the circularly polarized emission from \roa\
(bottom panel of Fig.~\ref{fig_spec}).
Finally, the presence of ranges of phases 
where the Stokes\,{\em V} emission at $\nu \geq 5.5$ GHz was undetected 
indicates that at these phases the magnetic axis of \roa\ is almost
perpendicularly oriented to the line of sight.
The above geometric condition is related to the nulls of the effective magnetic field curve.

The average radio spectrum of the total intensity (Stokes\,{\em I}) of \roa\ is given in the top panel of Fig.~\ref{fig_spec}.
Like in the cases of other hot magnetic stars well studied at radio wavelengths \citep{leto_etal17a,leto_etal18},
\roa's spectrum can be fitted using a power law with a quite flat spectral index 
($\approx 0.1$ within the spectral range 2.1--21.2 GHz).
The corresponding average radio spectral luminosity is $L_{\nu,\mathrm{rad}}\approx 1.8 \times 10^{17}$ erg s$^{-1}$ Hz$^{-1}$.

The radio spectral luminosity of \roa\ is about one order of magnitude smaller then the radio spectral luminosities of 
three similar hot magnetic stars: 
HD\,37479 \citep*{linsky_etal92}, HR\,7355 \citep{leto_etal17a}, and HR\,5907 \citep{leto_etal18},
which have radio spectral luminosities $\sim 10^{18}$ erg s$^{-1}$ Hz$^{-1}$ and polar field strengths $\sim 10^4$ G.
Comparing with CU\,Vir 
($L_{\nu,\mathrm{rad}}\approx 3 \times 10^{16}$ erg s$^{-1}$ Hz$^{-1}$, \citealp{leto_etal06}),
a cooler magnetic star (A0Vp) with a polar field strength of $3800$ G \citep{kochukhov_etal14},
the radio spectral luminosity of \roa\ is instead an order of magnitude higher.

To roughly estimate the radiative energy lost by \roa\ via incoherent gyro-synchrotron emission, 
we exploit the spectral behavior of the hot magnetic stars observed
at the millimeter wavelength range ($\nu > 100$ GHz).
The low level of detection rate \citep{leone_etal04}, combined with the flux drop at the high frequencies 
measured by ALMA \citep{leto_etal18},
makes us confident in constraining the upper limit of the gyro-synchrotron band at $\nu < 1000$ GHz.
Conservatively, we assume  the upper limit of the frequency band { is} equal to 1000 GHz.
Further, against the observational evidence,
we assume that the radio spectral luminosity remains flat within the radio frequency band 
where the incoherent gyro-synchrotron mechanism produces a detectable emission level.
Following the above assumptions,
within a frequency band $\sim 10^{12}$ Hz wide, the upper limit of the radio power of \roa\ is 
$L_{\mathrm{rad}}\approx 1.8 \times 10^{29}$ erg s$^{-1}$.

\subsection{Coherent radio emission}
\label{radio_coe}
\begin{figure}
\resizebox{\hsize}{!}{\includegraphics{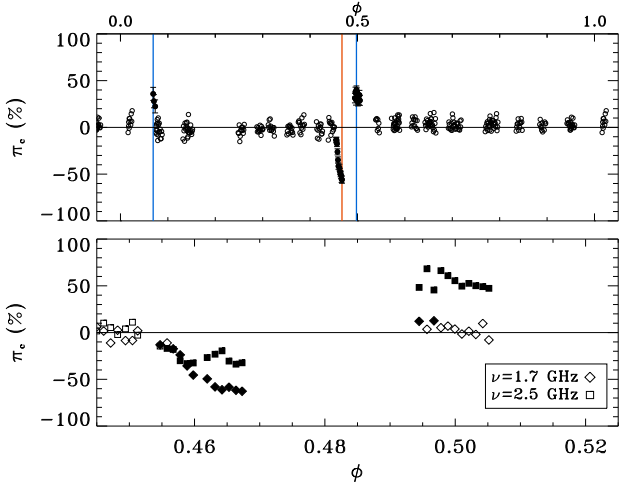}}
\caption{Top panel: rotational modulation of the fraction of the \roa\ circularly polarized emission measured at the L band.
The vertical lines locate the phases of the polarized emission maxima. Blue lines refer to the RCP polarization sense,
the red line to the LCP.
Bottom panel: spectral behavior of the \roa\ highly polarized emission close to $\phi \approx 0.5$.
Two sub-bands $\approx 1$ GHz wide, { centred} respectively at $\nu=1.7$  GHz (diamond symbol)
and 2.5 GHz (square symbol), were separately displayed.
The filled symbols (top and bottom panels) highlight the measurements above the $3 \sigma$ threshold.}
\label{fig_are_zoom}
\end{figure}

The L band rotational modulation of $\pi_{\mathrm c} \times 100$   
is shown in the top panel of Fig.~\ref{fig_are_zoom}.
The highly polarized transient detected at 2.1 GHz during the third observing run
(occurring at $\phi \approx 0.5$) shows a clear helicity reversal for the sense of circular polarization 
within a narrow range of phases ($\Delta \phi \approx 0.03$).
The measured percentage of $\pi_{\mathrm c}$ varies from a level of $\approx -60$ per cent
of LCP emission 
to $\approx +40$ per cent of RCP emission.
It is worth noting that there is also a hint of highly polarized emission, above the detection threshold, at $\phi \approx 0.07$.

The phase zoom of the polarized burst at  $\phi \approx 0.5$ is displayed in the bottom panel of Fig.~\ref{fig_are_zoom}.
Within the L band receiver bandpass, a clear spectral dependence of this strong polarized burst is evident.
In particular, the LCP emission is maximum at  $\nu=1.7$ GHz ($\pi_{\mathrm c} \approx -63$ per cent),
whereas the RCP maximum  ($\pi_{\mathrm c} \approx68$ per cent) was measured at $\nu=2.5$ GHz.
The ATCA  measurements were performed by cyclically varying the observing bands, which prevents a broader frequency study of this fast event.
Such a high level of low-frequency polarized emission cannot be explained as a canonical incoherent gyro-synchrotron emission,
Sec.~\ref{radio_incoe}.
The ATCA L~band observations of \roa\ 
have sampled the range of phases { centred} at $\phi \approx 0.5$ just once.
This prevents us from assessing whether the behavior is stable with time.

Such a low-frequency highly polarized emission,
constrained within a narrow range of stellar rotational phase, 
is indicative of a coherent emission mechanism and { has also been}
observed in other hot magnetic stars
\citep*{trigilio_etal00,das_etal18,leto_etal19,das_etal19a,das_etal19b}.
The elementary amplification mechanism for this type of coherent emission
is the Electron Cyclotron Maser (ECM) powered by an unstable electron energy distribution
\citep{wu_lee79, melrose_dulk82, winglee_pritchett86}.
It is worth { noting that the number of early type stars discovered as coherent sources is rapidly increasing.
This research field is fast progressing,
just a few months before the end of the year 2019 the detection of coherent
radio emission from HD\,35298 was reported \citep{das_etal19b}. 
Note that among this sample of stars} 
\roa\ is the hottest in which the ECM emission has so far been detected,
previously HD\,142990 ($\approx 18$ kK) was reported as the hottest star with ARE \citep{das_etal19a}.

\section{The ORM geometry of \roa}
\label{beff_curve}

The stellar rotation period of \roa\ derived in Sec.~\ref{sec:ephe} ($\approx 0.75$ days) 
is very close to the rotation period of another B2-type star: HD\,345439, $P_{\mathrm {rot}}\approx 0.77$ 
days \citep{wisniewski_etal15,hubrig_etal17}.
At the present time, \roa\ is
the third-most rapidly rotating magnetic B-type star yet discovered,
after HR\,5907 \citep{grunhut_etal12} and HR\,7355 \citep{rivinius_etal13},
both stars having rotation periods of $\approx 0.5$ days.

The fast rotation of \roa\ might produce rotational oblateness. 
We calculate the ratio between the polar and equatorial radii
following the procedure described in \citet{shultz_etal19b}.
We estimated $R_{\mathrm{p}}/R_{\mathrm{eq}}=0.86 \pm 0.02$,
which puts \roa\ between the cases of the two fast rotators: HR\,7355 ($R_{\mathrm{p}}/R_{\mathrm{eq}}=0.83$)
and HR\,5907 ($R_{\mathrm{p}}/R_{\mathrm{eq}}=0.88$)
\citep{grunhut_etal12,rivinius_etal13}.
{ Assuming in first approximation the value of the polar radius not depending by the stellar rotation \citep{maeder09},
using for $R_{\mathrm p}$ the value of
the stellar radius listed in Table~\ref{par_star},
we estimate $R_{\mathrm{eq}} \approx 5.2$ R$_{\odot}$.}
The equatorial radius is used to constrain the inclination of 
the rotation axis for \roa.
The measured projected rotation velocity and the rotation period
listed in Table~\ref{par_star}, allow us to derive an inclination angle $i \approx 35 ^{\circ}$. 

\begin{figure}
\resizebox{\hsize}{!}{\includegraphics{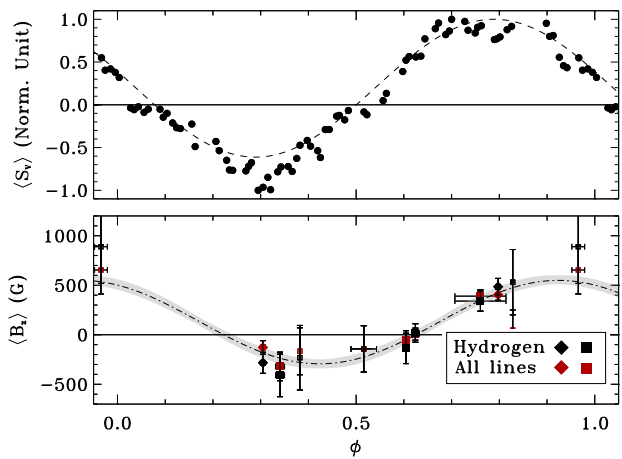}}
\caption{Top panel: average of the circularly polarized radio emission of \roa\ at $\nu\geq 5.5$ GHz.
Each single band dataset was normalized to the their extrema (positive/negative).
The normalized data, averaged within a phase bin 0.01 large, are displayed using the filled dots.
The dashed line is the sinusoidal fit of the data.
Bottom panel:  effective magnetic field curve of \roa.
The open symbols refer the \bz\ measurements obtained from lower quality spectra (quality threshold S/N $>1000$).
The gray area represents the envelope of the sinusoidal curves compatible with the errors of the data fit parameters.}
\label{fig_b}
\end{figure}

The effective magnetic field measurements of \roa, listed in Table~\ref{tab:Bfield},
displayed a sign reversal, indicative of a north-south magnetic hemisphere visibility switch.
In the ORM framework, the misalignment between magnetic and rotation axes (angle $\beta$)
is related to the ratio between the minimum and maximum effective magnetic field ($r$)
by the relation: $\tan \beta \tan i = (1-r)/(1+r)$  \citep{preston67}.
Using the sinusoidal fit of the phase folded \bz\ measurements of \roa\ (see bottom panel of Fig.~\ref{fig_b}),
we estimate $r=-0.55 \pm 0.1$, hence it follows that $\beta \approx 80^{\circ}$.

For a simple dipolar topology,
the polar magnetic field strength ($B_{\mathrm p}$) is related to the maximum measured effective magnetic field
by the relation \citep{schwarzschild50}:
\begin{displaymath}
B_{\mathrm p}= | \langle B_z\rangle (max)|  \frac{4(15-5u)}{15+u} \cos (i-\beta),
\end{displaymath}
where $u$ is the limb darkening coefficient (listed in Table~\ref{par_star}).
The $u$ parameter was computed and tabulated for different atmosphere models \citep{claret_etal11}.
Among the tabulated values,
we retrieved the vale of $u$ in the visual band corresponding to the stellar parameters of \roa.
The maximum value retrieved by the sinusoidal fit of the available data is $\langle B_z\rangle (max)=550 \pm 50$ G. 
Consequently, the polar field strength of \roa\ is { $B_{\mathrm p} \approx 2700$ G.}

\begin{figure*}
\resizebox{\hsize}{!}{\includegraphics{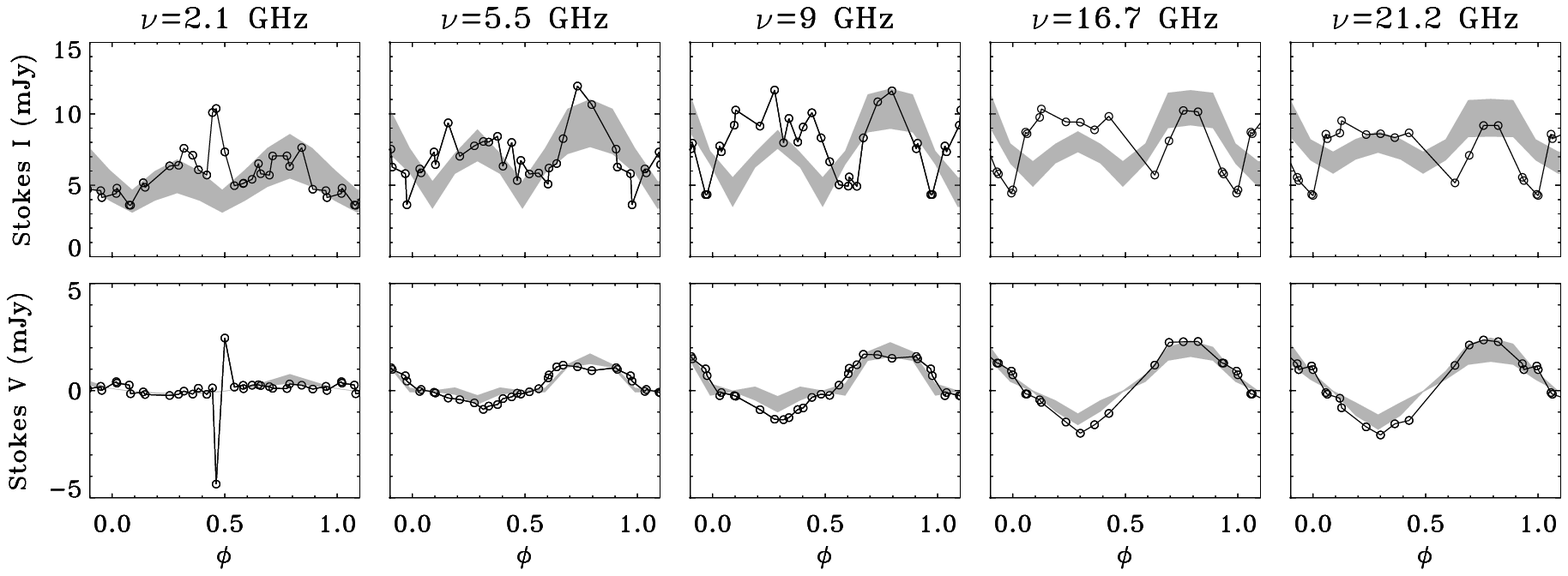}}
\caption{Comparison between synthetic and observed radio light curves of \roa.
Top panels: total intensity (Stokes\,$I$) corresponding to each observing band. 
Open circles are the values averaged 
over the time duration of the individual ATCA observing scans.
Bottom panels: circularly polarized intensity (Stokes\,$V$).
Grey areas represent the envelope of the synthetic radio light curves corresponding to each set of the model free parameters
able to produce simulations that are fairly in accordance with the observations.}
\label{simulaz_3d}
\end{figure*}

The derived ORM parameters (listed in Table~\ref{par_star}) allow us to calculate the
effective magnetic field curve following the method described in \citet{leto_etal16}.
The synthetic effective magnetic field curve is displayed in the bottom panel of Fig.~\ref{fig_b},
this curve is perfectly in accordance with the sinusoidal fit of the data,
thus validating our analysis of the stellar magnetic field strength and its geometry.

The ORM geometry was also compared with the behavior
{ of the Stokes\,{\em V} radio light curves.}
The Stokes\,{\em V} measurements at $\nu \geq 5.5$ GHz display a sinusoidal rotational modulation 
(see right panels of Fig.~\ref{l_fase}).
The amplitude variation of the Stokes~{\em V} light curves grows as the observing frequency increases
(see bottom panel of Fig.~\ref{fig_spec}). 
Then, to compare the measurements performed at different observing bands (frequency range 5.5--21.2 GHz),
each single band dataset was normalized to their extrema (positive/negative).
The normalized data, averaged within a phase bin of width 0.01, are displayed in the top panel of Fig.~\ref{fig_b},
with a simple sinusoidal fit superimposed.

The two observables reported in Fig.~\ref{fig_b} trace the stellar magnetic field at different heights.
The effective magnetic field is the average over the whole visible disk 
of the longitudinal components of the magnetic field vectors anchored to the stellar surface.
However, the circularly polarized radio emission is sensitive to the magnetic field topology that characterizes
the magnetospheric regions where radio emission mainly originates,
regions that are well above the stellar surface.
The higher { multipoles'} contributions decrease with radial distance more rapidly than the lower ones, 
{ thus,} moving outward from the stellar surface the magnetic field will be dominated by the simple dipole component.
The \bz\ measurements are sensitive to the magnetic topology at the stellar 
{ surface; if} higher multipoles are present, 
the dipole axis orientation that characterizes the magnetic topology at large distance
might deviate from that derived using the simple ORM geometry for fitting 
the true (non-dipolar?) magnetic topology measured at the stellar surface.
Hence, a phase shift between radio and \bz\ modulations is expected when the field is not a simple dipole.
Such behavior was clearly observed in the case of CU\,Vir \citep{kochukhov_etal14} and HD\,142301 \citep{leto_etal19}.
The comparison between the two curves pictured in Fig.~\ref{fig_b} shows
that a phase shift is also present in the case of \roa.
The extrema of the circularly polarized emission are in advance ($\Delta \phi \approx 0.1$) 
with respect to the stellar orientations 
{ at which the magnetic poles are more visible,}
suggesting that the simple magnetic dipole
is only a first approximation of the more complex magnetic field topology of \roa,
{ although  the relatively sparse rotational phase coverage in the \bz\ measurements
cannot definitively confirm it. 
}

\section{The magnetosphere of \roa}
\label{sec_magsph}

Even if the distribution of some chemical elements is anisotropic
(Sec~\ref{sec:ephe}),
the average chemical composition of \roa\ is quite similar to that of the Sun (Sec~\ref{spectra}).
We used the scaling relations from \citet{krticka14}
(derived using a fixed stellar radius corresponding to mean main-sequence stars),
which hold in the case of B type stars with solar chemical abundances,
to estimate the mass loss rate ($\dot M$) 
and wind terminal velocity ($v_{\infty}$) of \roa.
Adopting the effective
temperature listed in Table~\ref{par_star}
and correcting for the actual stellar radius,
assuming the luminosity dependence 
$\dot M \propto (L/{\mathrm L_{\odot}})^2$
\citep{krticka_kubat12},
we estimate $\dot M \approx 1.2\times10^{-10}$ M$_\odot$\,yr$^{-1}$
and $v_{\infty} \approx 1500$ km s$^{-1}$.  
The obtained values are similar to those found empirically for B-type stars \citep{oskinova_etal11}.

In the presence of a large-scale stellar magnetic field, the ionized
wind material cannot freely propagate.  At distances lower then the
Alfv\'{e}n radius, the ionized wind plasma is confined by the magnetic
field.  For a simple dipolar field topology, neglecting the stellar
rotation, $R_{\mathrm {A}}$ is related to the wind confinement parameter
$\eta_{\ast} = B^2_{\mathrm {eq}} R^2_{\ast} / \dot M v_{\infty}$
\citep{ud-doula_owocki02} (where $B_{\mathrm {eq}}$ is the magnetic
field strength at the stellar equator, which for a simple dipole is half
of the polar value) with $R_{\mathrm {A}} \propto \eta_{\ast}^{1/4}$
\citep*{ud-doula_etal08,ud-doula_etal14}.  Using the wind
parameters above with the stellar parameters in Table~\ref{par_star},
we estimate $R_{\mathrm A} \approx 20$ $R_{\ast}$.

However, \roa\ is also a fast rotator, and
rotation effects could alter the value of $R_{\mathrm A}$.  
Consequently, we can indirectly estimate the size of the middle magnetosphere
of \roa\ by modeling the multi-wavelength radio light curves, both for
the total intensity (Stokes\,$I$) and the circularly polarized emission
(Stokes\,$V$). 
The 3D-model of the gyro-synchrotron emission from a
dipole-shaped stellar magnetosphere \citep{trigilio_etal04,leto_etal06}
was applied using the polar field strength and the ORM geometry of \roa\
(parameters listed in Table~\ref{par_star}).  

The model has been developed under the hypothesis of spherical symmetry.
The simulations were performed using the average stellar radius of \roa\
($R_{\ast}$, listed in Table~\ref{par_star}).  To reproduce the observed
shape of the multi-wavelength radio light curves of \roa, the Alfv\'{e}n
radius was varied in the range 5--25 $R_{\ast}$, with a simulation
step of 1 $R_{\ast}$.  The relativistic electrons were assumed power
law energy distributed: $N(E) \propto E^{-\delta}$.  The simulations
were performed using two different values of the spectral indices
$\delta=2$ and $\delta=2.5$, in accordance with the results retrieved
by the simulations of the radio emission of other hot magnetic stars
\citep{leto_etal06,leto_etal17a,leto_etal18}.  The equatorial thickness
of the middle-magnetosphere ($l$) was varied between 10 per cent and 100 per cent of
$R_{\mathrm A}$.  The relativistic electron density ($n_{\mathrm r}$),
{ which is} responsible for the non-thermal radio emission, was varied in
the range $10^{2}$--$10^{5}$ cm$^{-3}$, { and the} adopted simulation step was
$\Delta \log n_{\mathrm r} \approx 0.1$.

The temperature and density of the thermal plasma trapped within the inner-magnetosphere 
are functions of the radial distance.
In accordance with the MCWS model, 
the temperature linearly increases and the density linearly decreases moving outward.
The temperature at the stellar surface is assumed equal to $T_{\mathrm {eff}}$, whereas the corresponding thermal electrons density ($n_{0}$)
was varied between $10^7$ and $10^{10}$ cm$^{-3}$, simulation step $\Delta \log n_{0}=0.5$.

Even if the simple dipole-like topology is a first approximation of the true magnetic field topology of \roa, as discussed in Sec.~\ref{beff_curve},
among the explored sets of model free parameters, we found some combinations that are able to 
simulate incoherent emission that is in agreement 
{ with the radio spectrum as well as the modulation of the radio light curves of \roa.}
The synthetic radio light curves that better resemble the observed ones are
displayed in Fig.~\ref{simulaz_3d}. Top panels of the figure refer to the total intensity (Stokes\,$I$),
bottom panels to the circularly polarized emission (Stokes\,$V$).
Looking at the figure, it is evident that
the Stokes\,$V$ simulations are well in accordance with the observations, whereas some discrepancies
between observations and simulations 
are evident for the Stokes\,$I$.
The circularly polarized emission is sensitive to the ordered magnetic field, whereas 
also regions where the magnetic field is strongly anisotropic contributes to the total intensity.
This is further evidence 
that the overall magnetic field topology of \roa\ deviates from a simple dipole. 

Using models simulations, we constrain several physical
parameters of the \roa\ magnetosphere.  
First, 
the simulations
predict $R_{\mathrm A}$ in the range 8--12 $R_\ast$.
Both values of 
$\delta$ were found to produce synthetic radio light curves similar
to the observed ones.  The two parameters $l$ and $n_{\mathrm r}$ are
degenerate, hence we can only retrieve the equatorial relativistic column
density at the Alfv\'en radius.  For $\delta=2$, $n_{\mathrm
r} \times l = 10^{14.8 \pm 0.1}$ cm$^{-2}$; for $\delta=2.5$,
$n_{\mathrm r} \times l = 10^{15.5 \pm 0.2}$ cm$^{-2}$.  Finally, the
effects of thermal free-free absorption are able to reproduce the
observed rotationally modulated amplitudes, with trapped thermal
electrons having an average energy $kT \approx 0.01$ keV and density $n_{0}$
at the stellar surface in the range $10^{8.5}$--$10^{9}$ cm$^{-3}$.

The model also permits an estimate of the wind mass-loss rate
by using the radio { measurements} only.  
In fact, the equatorial region of the stellar magnetosphere 
where the thermal plasma opens the magnetic
field lines can be estimated from equating the magnetic energy density of the
dipolar stellar field with the wind energy density, including the influence
of the centrifugal component from stellar rotation
(for details see \citealp{trigilio_etal04}).
The size of \roa's magnetosphere { enables us} to reproduce the observed radio measurements 
{ with a mass-loss rate} in the range 
$2.6\times10^{-10}$--$2.1\times10^{-9}$ M$_\odot$\,yr$^{-1}$, 
%
the corresponding values of $\eta_{\ast}$ would become 
$\approx 9.1 \times 10^3$--$7.3 \times 10^4$.
%
The wind mass-loss rate estimated using the scaling relation
of the B-type star's wind is lower than the lower limit 
derived by modeling the radio emission of \roa.
As discussed by \citet{krticka14},
deviations from the typical Solar abundance could significantly
modify the mass-loss rate of the radiatively driven wind from the B type stars.  
We anticipate that the chemical composition of \roa\ may slightly deviate from that of the Sun.

The wind power $\frac{1}{2} \dot M v_{\infty}^2$ corresponding to
the mass loss rate of \roa, estimated by the radio emission modeling,
holds in the range $10^{32}$--$10^{33}$ erg s$^{-1}$.  But, due to
the presence of the large scale stellar magnetic field, the wind
plasma can escape only from the polar caps, where the magnetic
field lines are open. Then the effective (or ``actual'')
mass-loss rate of \roa\ will be 
$\dot{M}_{\mathrm{act}}=1.4\times10^{-10}$--$1\times10^{-11}$ M$_\odot$\,yr$^{-1}$.
%

The mass lost via the wind also transports angular momentum and leads to
magnetic braking. The corresponding spin-down time ($\tau_{\mathrm{spin}}$)
of \roa\ can be estimated using the relation given by
\citet{ud-doula_etal09}:

\begin{displaymath}
\tau_{\mathrm{spin}} \approx 110 \frac{0.1}{B_{\mathrm p}/{\mathrm {kG}}} \frac{M/{\mathrm M}_{\odot}}{R_{\ast}/{\mathrm R_{\odot}}} \sqrt{\frac{v_{\infty} / (10^3 \mathrm {km \, s^{-1}})}{\dot M / (10^{-9}\mathrm{M_\odot \, yr^{-1})}}}~{\mathrm {(Myr),}}
\end{displaymath}

\noindent which produces a magnetic braking timescale of \roa\ in
the range  $\approx 24$--90 Myr. 
We can estimate an upper limit to
the spin-down age ($t_{\mathrm s}$) of \roa\ since arriving on the
zero age main sequence (ZAMS).  The age relation is $t_{\mathrm
{s,\,max}}= \tau_{\mathrm{spin}} \ln (1/W)$, where the critical
rotation parameter $W=v_{\mathrm{eq}} / v_{\mathrm{c}}$ is the ratio
of the stellar equatorial rotation speed $v_{\mathrm{eq}}=\omega
R_{\mathrm {eq}}$ (with $\omega=2\pi / P_{\mathrm{rot}}$ angular
velocity) and the critical speed of rotation $v_{\mathrm{c}}=\sqrt{G
M_{\ast} / R_{\mathrm {eq}}}$ (where $G$ is the gravitational
constant).  The relation was derived by \citet{petit_etal13} under
the initial condition that the star rotates at critical speed upon
arriving on the ZAMS ($W_0=1$).  For \roa, $W \approx 0.64$, 
and $t_{\mathrm {s,\,max}}\approx11$--40 Myr, 
which is longer than the age of the $\rho$\,Oph group (5--10 Myr; \citealp{pillitteri_etal16}).
For \roa, an older age of 15.3~Myr was determined
\citep{pillitteri_etal18}, that is still compatible with the range of
spin-down ages above estimated.

Finally, we estimated the Kepler co-rotation radius ($R_{\mathrm{K}}$)
of \roa.  $R_{\mathrm{K}}$ is the distance from the center of the
star where the centrifugal acceleration equates the gravitational
one, then, this parameter can be retrieved from the relation:
$R_{\mathrm{K}} \omega^2 = G M_{\ast} / R_{\mathrm{K}}^2$.  Using
again the rotation parameter $W$, it is possible to write the scaling
relation: $R_{\mathrm{K}} = W^{-2/3} R_{\mathrm {eq}}$, that { allows us} 
to derive $R_{K}$ as a function of $W$.  Using the stellar parameters
listed in Table~\ref{par_star} we derived 
$R_{\mathrm{K}}\approx1.35$ $R_{\mathrm {eq}} \approx 1.6$ $R_{\ast}$.  
As a result of the radio
emission modeling, we estimated the average value of the equatorial
stellar Alfv\'{e}n radius of \roa\ ($R_{\mathrm{A}}\approx 10$
$R_{\ast}$).  Comparing the values of these two radii, it follows
that $R_{\mathrm{K}} \ll R_{\mathrm{A}}$, classifying \roa\ as a
star with a centrifugal magnetosphere. 
We also find that the dimensionless parameter 
$\log(R_{\mathrm A} / R_{\mathrm K})$ is $\approx 0.8$, placing \roa\ just at  the boundary
of where magnetic early-type stars are H$\alpha$ emitters, which
are characterized by $\log(R_{\mathrm A} / R_{\mathrm K}) \gtrapprox
0.8$ \citep{petit_etal13,shultz_etal19b}.  

\section{The X-ray emission of \roa}
\label{sec_x_ray}
Two X-ray bursts from \roa, separated by $\approx 1.2$ days,
were detected by \XMM\ during observations performed in 2016.
The basal (quiescent) X-ray emission was instead observed almost steady and well in accordance 
with the \XMM\ measurements performed in 2013 \citep{pillitteri_etal14}.
The two observing epochs are separated by more than 2.5 yrs,
suggesting a steady mechanism of the \roa\ X-ray quiescent emission.

The average quiescent X-ray flux of \roa\ is $\approx 1.7\times
10^{-12}$ erg s$^{-1}$ cm$^{-2}$ corresponding to an X-ray luminosity
of $\approx 4 \times 10^{30}$ erg s$^{-1}$.  The MCWS model
predicts an X-ray luminosity in the
range $10^{31}$--$10^{32}$ erg s$^{-1}$, derived using the scaling
law: $L_{\mathrm X} \propto \dot{M} v_{\infty} B_{\mathrm p}^{0.4}$
(the range of wind mass loss rate was retrieved in
Sec.~\ref{sec_magsph} and listed in Table~\ref{par_star}), which
is significantly higher than the measured quiescent emission level.
The XADM model predicts 5 times lower X-ray luminosity, 
in this case the measured X-ray
quiescent level lies within the expected theoretical range 
($2 \times 10^{30}$--$2 \times 10^{31}$ erg s$^{-1}$).
During the two X-ray bursts observed in 2016, \roa\ reached fluxes of $\approx 4
\times 10^{-12}$ and $\approx 8 \times 10^{-12}$ erg s$^{-1}$
cm$^{-2}$, respectively.  The X-ray luminosities of these two events
were $L_{\mathrm X}\approx 0.9 \times 10^{31}$ and $\approx 1.9
\times 10^{31}$ erg s$^{-1}$, which are compatible with
the range predicted by the XADM model, even if close to the upper boundary.  
Interestingly, \cite{naze_etal14} found that stars with large centrifugal magnetospheres can be overluminous in
X-rays as compared to the predictions of XADM.  Consequently,
the high X-ray luminosity of \roa\ further supports the
presence of a centrifugally supported magnetosphere for this star.

The ratios between the X-ray luminosity and the radio (Sec.~\ref{radio_incoe}) spectral luminosity
($L_{\mathrm X} / L_{\nu,\mathrm {rad}}$) are: 
$\approx 10^{13.3}$ Hz during the quiescent X-rays emission; 
$10^{13.7}$ Hz during the first X-ray pulse; 
$10^{14}$ Hz during the second and strongest pulse. 
It is worth { noting} that the above ratios between the X-ray and radio luminosities 
of \roa\ violate the empiric G\"{u}edel-Benz relation: $L_{\mathrm X} / L_{\nu,\mathrm {rad}} \approx 10^{15.5}$ Hz
\citep{guedel_benz93,benz_guedel94},
that is valid for main sequence stars ranging from the F to the early M spectral types.
The X-ray behavior of \roa\ is quite intriguing,
in fact, \roa\ comes closest to the G\"{u}edel-Benz  relation during its X-ray bright states.

\begin{table}
\caption[ ]{Parameters used to fit the X-ray spectra of \roa\  acquired during the faint and the bright states \citep{pillitteri_etal17}.
Brackets provide the related uncertainties.
The spectral fits were performed assuming two models: three thermal components, 
and two thermal components plus power-law.}
\label{par_xray}
\footnotesize
\begin{tabular}[]{lccc}
\hline
 & Quiescient  &1$^{\mathrm{st}}$ burst &2$^{\mathrm{nd}}$ burst \\
\hline
\multicolumn{4}{l}{3T thermal model}\\
\hline
$kT_1$  {\scriptsize {(keV)}}                                       &0.37{\scriptsize(0.05)}        &0.19{\scriptsize(0.09)}         &0.96{\scriptsize(0.06)}   \\
$EM_1$ {\scriptsize {($10^{53}$ cm$^{-3}$)}}        &0.52{\scriptsize(0.08)}           &1.5{\scriptsize(1)}       &4.0{\scriptsize(0.6)}   \\
\smallskip
\smallskip
Flux$_1$  {\scriptsize {($10^{-12}$ erg s$^{-1}$ cm$^{-2}$)}}   &0.21{\scriptsize(0.04)}     &0.4{\scriptsize(0.25)}       &1.9{\scriptsize(0.4)}   \\

$kT_2$ {\scriptsize {(keV)}}                                                                 &0.91{\scriptsize(0.03)}      &0.9{\scriptsize(0.1)}       &3.3{\scriptsize(0.3)}    \\
$EM_2$ {\scriptsize {($10^{53}$ cm$^{-3}$)}}                                 &1.17{\scriptsize(0.08)}     &1.5{\scriptsize(0.3)}       &1.1{\scriptsize(0.2)}     \\
\smallskip
\smallskip
Flux$_2$ {\scriptsize {($10^{-12}$ erg s$^{-1}$ cm$^{-2}$)}}     &0.63{\scriptsize(0.04)}          &0.8{\scriptsize(0.2)}       &2.2{\scriptsize(0.3)}   \\

$kT_3$ {\scriptsize {(keV)}}                                                                 &2.2{\scriptsize(0.1)}      &3.5{\scriptsize(0.4)}         &4.1{\scriptsize(0.4)}    \\
$EM_3$ {\scriptsize {($10^{53}$ cm$^{-3}$)}}                                 &1.9{\scriptsize(0.1)}     &4.9{\scriptsize(0.4)}      &8.1{\scriptsize(0.4)}     \\
\smallskip
\smallskip
Flux$_3$ {\scriptsize {($10^{-12}$ erg s$^{-1}$ cm$^{-2}$)}}     &0.89{\scriptsize(0.04)}      &2.8{\scriptsize(0.2)}       &3.5{\scriptsize(0.6)}   \\

Reduced $\chi^2$                                                         &1.94     &1.10   &1.27  \\
d. o. f.                                                                               & 94          & 38          & 41     \\
\hline

\multicolumn{4}{l}{2T thermal $+$ power-law model}\\

\hline
$kT_1$ {\scriptsize {(keV)}}                                                                                              &0.82{\scriptsize(0.02)}  &0.90{\scriptsize(0.08)}  &1.0{\scriptsize(0.1)}   \\
$EM_1$ {\scriptsize {($10^{53}$ cm$^{-3}$)}}                                                            &1.22{\scriptsize(0.05)}  &1.2{\scriptsize(0.4)}  &3{\scriptsize(1)}   \\
\smallskip
\smallskip
Flux$_1$ {\scriptsize {($10^{-12}$ erg s$^{-1}$ cm$^{-2}$)}}                           &0.68{\scriptsize(0.03)}    &1.2{\scriptsize(0.8)}      &1.5{\scriptsize(0.5)}   \\

$kT_2$ {\scriptsize {(keV)}}                                                                                      &2.0{\scriptsize(0.1)}  &2.8{\scriptsize(0.4)}  &1.9{\scriptsize(0.8)}   \\
$EM_2$ {\scriptsize {($10^{53}$ cm$^{-3}$)}}                                                           &1.7{\scriptsize(0.1)}  &0.02{\scriptsize(0.01)}      &8{\scriptsize(4)}   \\
\smallskip
\smallskip
Flux$_{{2}}$ {\scriptsize {($10^{-12}$ erg s$^{-1}$ cm$^{-2}$)}}                          &0.81{\scriptsize(0.07)}        &0.7{\scriptsize(0.2)}    &3.8{\scriptsize(0.7)}   \\

$K$  {\scriptsize {($10^{-4}$ ph keV$^{-1}$ s$^{-1}$ cm$^{-2}$ at 1 keV)}}        &0.5{\scriptsize(0.2)}           &7{\scriptsize(2)}       &3{\scriptsize(1)}              \\
$\alpha$                                                                                                                           &2.4{\scriptsize(0.2)}           &2.2{\scriptsize(0.2)}  &1.3{\scriptsize(0.2)}   \\
\smallskip
\smallskip

Flux$_{\mathrm{pow}}$ {\scriptsize {($10^{-12}$ erg s$^{-1}$ cm$^{-2}$)}}  &0.25{\scriptsize(0.1)}              &2{\scriptsize(1)}       &2.3{\scriptsize(0.5)}   \\

Reduced $\chi^2$                                                                          &2.09        &1.36    & 1.14  \\
d. o. f.                                                                                               &  94         &  38         & 37     \\

\hline
\end{tabular}
\end{table}

The radio and the X-ray emission that characterize the stellar
magnetic activity of the late-type stars and the Sun is well understood.
In the case of Solar flares, the energy release occurs { at the top}
of the coronal magnetic loops, where the local plasma
is accelerated { to the} relativistic regime. These non-thermal electrons
impacting with the photosphere produce hard X-rays at the magnetic
loop footprints by means of a thick-target bremsstrahlung emission
mechanism, whereas a softer X-ray component is radiated by the
evaporation of chromospheric plasma that rises to the top of the
flaring loop (see \citealp{aschwanden02} and references therein).

The spectral index ($\alpha$) of the non-thermal X-ray photons
produced by the thick-target bremsstrahlung emission mechanism
is related to the spectral index ($\delta$) of the relativistic electrons impacting with the surface
by the simple relation: $\alpha=\delta -1$ \citep{brown71}.
The relativistic electrons responsible for the X-ray emission from the loop footprints
are the same non-thermal electrons that fills the flaring magnetic loops and 
that radiates at the radio regime by the incoherent gyro synchrotron emission mechanism.

For the hot magnetic stars, the energy release responsible for the
non-thermal plasma occurs in a magneto-disc located just outside
the equatorial Alfv\'{e}n radius, as in the magnetosphere of Jupiter
\citep{nichols11}. Here the acceleration process involves the whole
stellar magnetosphere, then, the corresponding magnetic footprints
are shaped like annular rings around the magnetic poles, similar
to the auroral rings observed in the magnetized planets of the solar
system \citep{badman_etal15}.  
The simulations performed in
Sec.~\ref{sec_magsph} allowed us to confirm that the radio emission
from \roa\ is produced by the gyro-synchrotron emission mechanism
from a population of non-thermal electrons power-law energy distributed
with spectral index $\delta=2$--2.5.  In order to identify evidence for 
non-thermal electrons based on the X-ray emission,
we re-analyzed X-ray spectra in quiescence and in outburst, with
the goal of checking for the presence of a non-thermal spectral
component.  Two distinct modeling approaches have been followed:
(1) a purely thermal ({\sc apec}) three component (3T) model, and
(2) a 2T thermal model plus non-thermal component, with $A(E) = K
E^{-\alpha}$, where $A$ is the specific photon flux.  For the ISM
hydrogen column density and abundance of the heavy elements, the
same values adopted by \citet{pillitteri_etal17} have been used.
The corresponding fit parameters are listed in Table~\ref{par_xray}.
The new spectral analysis indicates that the pure 3T thermal model
produces better quality fits of the X-ray spectra acquired during
the quiescent state and during the first burst, whereas for the
spectrum of the second burst the model including a non-thermal
component has a better quality { fit.}

Even if the $\chi^2$ statistic cannot clearly support one of the two proposed cases
(pure multicomponent thermal model vs thermal plus non-thermal model),
the possible existence of non-thermal X-ray photons within { \roa's spectra} 
can be related to the presence of the non-thermal electrons
responsible for incoherent gyro-synchrotron radio emission. 
The coincidence of the spectral index retrieved by the  
fit of the X-ray spectrum acquired during the second burst ($\alpha=1.3$),
with the value expected from the thick-target bremsstrahlung emission mechanism ($\alpha=1$--1.5)
could by a clue of the impact of the non-thermal electrons 
with the surface annular rings around the magnetic poles of \roa.
This might be a stellar analogue to the X-ray auroral emission
from Jupiter \citep{branduardi-raymont_etal07,branduardi-raymont_etal08}.

As discussed above, \roa\ displays considerable X-ray variability.
Following { the RFHD simulation approach \citep*{townsend_etal07}}
the X-ray emission
{ from hot magnetic stars is expected to be fairly steady.}
In fact the X-rays are produced far from the star from optically thin plasma.
On the other hand, if there is a significant contribution 
of non-thermal X-rays due to thick-target bremsstrahlung emission,
rotational variability of \roa's X-ray emission is expected. 
The source { regions} of such non-thermal X-rays { components are} likely shaped 
as bi-dimensional annular rings { centred} on the magnetic poles,
{ so} the X-ray emission level will { expected to be sensitive} to the visible geometrical area of the polar rings.

To explain the { \roa's observed X-ray variability,}
we studied the geometrical modulation of the visible area of the polar rings,
which are { the proposed sites} of the X-ray auroral emission, and
which could make a significant contribution to the total budget of the X-ray emission from \roa.
For an ORM, the visible area of the polar rings is a function of the stellar rotation.
To estimate the stellar rotation effect, we sampled the volume of the star using a cartesian grid.
The visible areas of all the elements of the stellar surface lying within a fixed range of magnetic latitudes ($\Lambda$)
have been added together. This was repeated varying the stellar rotational phase.
The above procedure was performed adapting the procedure developed to calculate \bz\ of an ORM as a function of the stellar 
rotation (see \citealp{leto_etal16} for details).

\begin{figure}
\resizebox{\hsize}{!}{\includegraphics{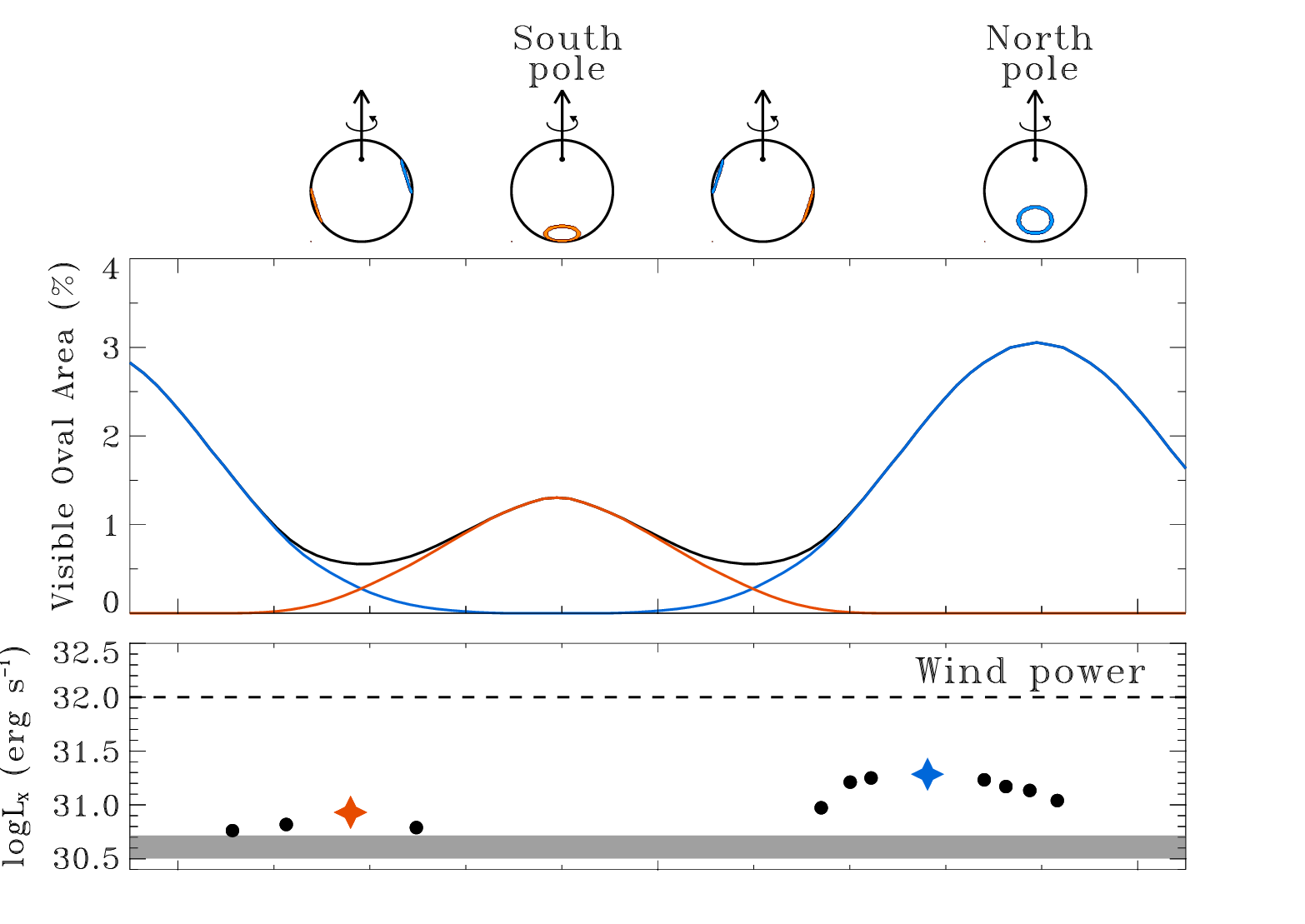}}
\resizebox{\hsize}{!}{\includegraphics{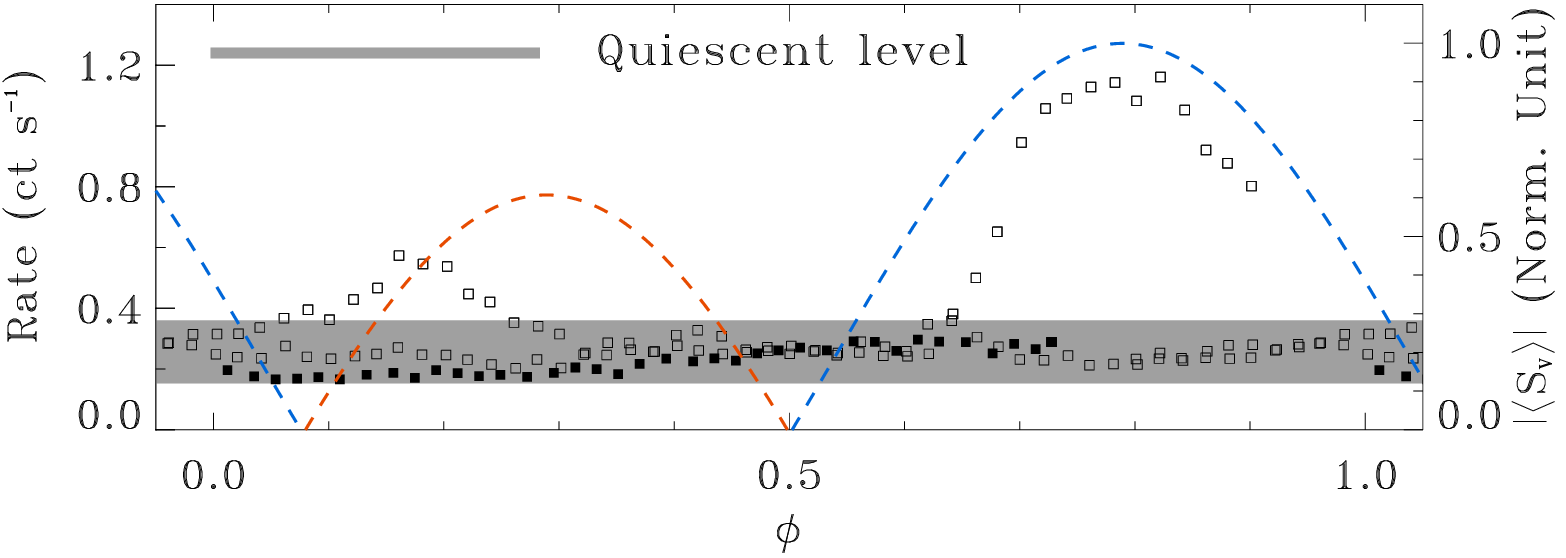}}
\caption{Top panel: fraction of the oval area visible from Earth during { \roa's} rotation.
The red line refers to the area of the southern oval, the blue line to the northern,
the black line is the sum of the contributions from { both ovals.}
The cartoons show the visibility of X-ray auroral rings
for four characteristic orientations of \roa.
The southern X-ray auroral ring is marked in red, the northern in blue.
Middle panel: measured X-ray luminosity during the bright states of \roa\ observed during { 2016.
The data} have been averaged over time (see \citealp{pillitteri_etal17}).
The X-ray data are phase folded using the period reported in Eq.~\ref{effemeridi}.
The red (1$^{\mathrm{st}}$ burst) and blue (2$^{\mathrm{nd}}$ burst) stars refer 
to the X-ray measurements used to perform the spectral analysis discussed in the text.
Bottom panel: light curve of the X-ray count rate phase folded using the same ephemeris. 
The X-ray measurements were averaged within phase bins of 0.02.
The 2013 observations \citep{pillitteri_etal14} are marked using the filled squares. 
The open squares refer to the 2016 observations \citep{pillitteri_etal17}.
The grey lines pictured in the bottom and middle panels locate the quiescent level of the X-ray emission.
The sinusoidal fit of the radio light curve of the incoherent circularly polarized emission in absolute value
is also reported in the bottom panel (red dashed line refers to the $\langle S_{\mathrm v}\rangle < 0$, blue to the $\langle S_{\mathrm v}\rangle  > 0$).}
\label{fig_model_x}
\end{figure}

The magnetic latitude of the polar ring is defined by the magnetic field line that crosses the magnetic equatorial plane 
at the distance equal to $R_{\mathrm{A}}$. For a simple magnetic dipole, the polar equation of the magnetic field line
is: $R= L \cos^2 \Lambda$  (where $L$ is the shell parameter, 
equal to the radial distance of the point where the magnetic field line crosses the magnetic equator).
Using the magnetosphere parameters given in Sec.~\ref{sec_magsph}, 
the $L$-shell parameter that locates the inner boundary of the middle-magnetosphere of \roa\  is 
in the range 8--12~$R_{\ast}$.
The corresponding magnetic latitudes at the stellar surface are $\Lambda \approx 69^{\circ}$ and $\Lambda \approx 73^{\circ}$.
Adopting the ORM of \roa,
we calculated the visible area of the northern and the southern polar rings 
with $\Lambda$ in the range $69^{\circ}$--$73^{\circ}$.
The fractional area of the visible auroral rings (with respect to the whole stellar disc)
is shown in the top panel of Fig.~\ref{fig_model_x} as a function of the stellar rotational phase.
The visible areas
of the two polar rings show a clear rotational modulation.
In particular, the geometry of \roa\ makes the northern ring { more visible}. 
In the top of Fig.~\ref{fig_model_x} are also pictured the cartoons showing the 
visible polar rings related to four particular stellar orientations, corresponding to 
the effective magnetic nulls and the \bz\ extrema.

We use the new ephemeris (Eq.~\ref{effemeridi}) to phase-fold the 2016 X-ray measurements.
The X-ray luminosities higher than the quiescent level are pictured in the middle panel of Fig.~\ref{fig_model_x},
the X-ray measurements of \roa\ acquired during the two brighter states,
and used for the spectral fitting process { discussed above},
are marked by using the { coloured} symbols.
The lower limit of the wind power produced by \roa\ is also reported
(horizontal dashed line in the middle panel of Fig.~\ref{fig_model_x}).
For comparison, the phase-folded X-ray fluxes (in counts of photons per second)
of \roa\ measured during the years 2013 \citep{pillitteri_etal14} and 2016 \citep{pillitteri_etal17}
are pictured in the lower panel of  Fig.~\ref{fig_model_x}.

Comparison between the geometrical simulation and the observations
reveals that the second and stronger event of enhanced X-ray
emission from \roa\ occurs at a phase where only the northern
polar ring is visible.  The first and weaker event is instead related
to a stellar orientation that makes visible almost the same fractional
area from both rings.  
Our geometrical analysis suggests that
significant fractions of the polar rings are always visible, hence,
the coincidence between the non-thermal X-ray emission and the
maximum polar ring area visibility seems not to be a stringent
condition for detecting non-thermal auroral X-rays.

The very fact that the X-ray light curve of \roa\ is rotationally modulated 
means that a significant fraction of X-rays 
comes from magnetospheric regions close to the star.
As discussed by \citet{pillitteri_etal17},
the bulk of the X-ray emission during the quiescent state
originates from plasma at a temperature in the range $\approx 10$--25 MK.
{ A further} contribution of hotter plasma (temperature $\approx 40$--50 MK) is instead needed
to reproduce the X-ray spectra acquired during the two bursts.
Following the non-thermal modeling approach,
the spectral contribution of this hot plasma is replaced by the non-thermal X-ray photons,
but the presence of thermal plasma, at a few tens of MK, is still confirmed.

The main source of the X-ray emission from hot magnetic stars 
is the kinetic energy of the colliding wind streams from the two opposite hemispheres \citep{babel_montmerle97,ud-doula_etal14}.
{ The Rankine-Hugoniot condition for a strong shock is} 
$T \approx 14 (v_{\mathrm{wind}}/10^3 {\mathrm{km~s}^{-1}})^2$ MK. 
Hence, for the maximum wind velocity $v_{\infty}=1500$ km s$^{-1}$,
the shocked plasma can reach a temperature of a few tens of MK, 
in accordance with the values { estimated above} by both modeling approaches.
However, for the second and higher peak, the \XMM\ spectrum shows the Fe\,{\sc xxv} line at $\lambda=1.8$~\AA~(6.7 keV)
that is surely associated with plasma temperatures in excess 
of 5.4 keV ($T \geq 62.7$ MK) and
cannot be explained by a power-law component. 
Thus a model describing the X-ray spectral features observed during this 
``bright state" of \roa\ requires the presence of a very hot thermal plasma,
as also suggested by the high temperature of the hottest thermal component of the pure thermal model (Table~\ref{par_xray}).

As discussed above, the MCWS model 
is unable to explain the temperature required to produce the 6.7 keV line.
{ Nevertheless, RFHD modeling of} the magnetically confined wind
predicts the possible existence of { highly energetic} plasma (temperature level $\approx 100$ MK)
within the { magnetosphere}. 
The RFHD simulation approach showed that such hot magnetospheric regions are located close to the star, 
where both hard and soft X-rays can be produced \citep*{townsend_etal07}.
Depending on the stellar geometry, these very hot plasma regions could be eclipsed as the star rotates, 
leading to a significant rotational modulation of the stellar X-ray emission.

On the other hand, the origin of such a hot thermal plasma component
could be also explained as a secondary effect of the non-thermal
particle bombardment of the stellar surface.  Not only non-thermal
X-rays are produced, but plasma evaporation can also result.  As
in the case of solar flares, evaporated plasma can fill magnetic coronal loops. 
This seems to be a reasonable possibility in the case of \roa.  
In fact, the comparison between the X-ray and radio light curves, shown in the bottom panel
of Fig.~\ref{fig_model_x}, indicates that both observables are
spatially related. The observational evidence suggests that both the
radio and the X-rays could arise from magnetospheric regions
that are mostly spatially coincident, and so the visibility of the
magnetospheric regions above the poles may also be a favorable
geometry for detecting thermal auroral X-rays.

\section{The auroral radio emission of \roa}
\label{sec_are}
The non-thermal electrons responsible for the incoherent gyro-synchrotron radio emission of \roa\
originate in acceleration regions located far from the star. 
Among the fast electrons propagating within the magnetosphere, 
only those with very low pitch angles
(angles between the electron velocity and the magnetic vector)
can impact on the stellar surface.
As a consequence, the reflected electron population (magnetic mirroring effect, \citealp{jackson62}) 
will be deprived of low pitch-angle electrons.
This condition promotes the development of the loss-cone unstable electron energy distribution,
that pumps the Electron Cyclotron Maser \citep{wu_lee79, melrose_dulk82}.

The ECM coherent emission mechanism
amplifies the radiation at frequencies close 
to the first few harmonics of the local gyro-frequency ($\nu_{\mathrm B} =2.8 \times 10^{-3} B/{\mathrm G}$ GHz).
But,
traveling trough the more external layers, where the local gyro-frequency is equal to the second harmonic of the amplified frequency,
the electromagnetic waves amplified at the fundamental harmonic
will be very likely suppressed by the gyromagnetic absorption effect \citep{melrose_dulk82}.
The elementary ECM emitting process produces amplified radiation constrained
within a very thin hollow-cone (axis coinciding with the local magnetic field vector) of large half aperture $\theta$ ($\approx 90^{\circ}$).
The hollow-cone thickness ($\Delta \theta$) and aperture 
are related to the electron velocity $v$ by the following relations: 
$\Delta \theta \approx v/c$;
$\cos \theta \approx v/c$ \citep{melrose_dulk82}, 
where $c$ is the speed of light.

Highly polarized amplified radio emission, powered by the coherent
ECM, is commonly detected from the magnetized
planets of the solar system and related to their planetary aurorae
\citep{zarka98}, or planetary ARE.  
The overall beam pattern of the ARE from a real source is the envelope
of the elementary ECM sources.  In the case of the ARE generated
in a thin magnetospheric cavity (i.e., laminar source region), the
escaping amplified radiation is confined within a narrow beam
tangentially directed along the cavity walls \citep{louarn_lequeau96}.

The observed highly polarized pulses from hot magnetic stars
are the stellar analogue of planetary ARE.  Following the above
``tangent plane beaming model'', stellar ARE will be seen in
the form of highly polarized radio pulses, as a consequence of
stellar rotation, making the stars where this coherent phenomenon
occurs similar to radio lighthouses.

As discussed above, the frequency of the ECM is directly related to the local magnetic field strength.
For a simple dipolar topology, the magnetic field strength
decreases outwards as $R^{-3}$, where $R$ is the radial distance from the star.
Thus for unstable electrons constrained within a thin magnetic shell,
the ECM source regions are rings located at different heights above the magnetic poles.
The observed large bandwidth of the ECM is a consequence of the wide range of 
magnetic field strengths covered by the regions where the maser amplification occurs.
The observed maser emission is the overlap of the narrow-band emission produced in magnetospheric layers 
at different local field strengths.
The lower frequencies of the ECM arise from farther regions, where the field is weaker,
whereas higher frequencies are generated close to the star. 
If the ECM is generated at the second harmonic of $\nu_{\mathrm B}$, 
the rings where { \roa's coherent} emission at the L-band arises are located in the range 0.82--1.07 
stellar radii above the poles,
(the adopted stellar radius and polar magnetic field strength are listed in Table~\ref{par_star}).
The first harmonic instead originates from layers located 0.44--0.64 $R_{\ast}$ 
above the surface.

The ECM beam forms a large angle $\theta$ ($\approx 90^{\circ}$)
with respect to the local magnetic field vector.
Then, assuming an absolutely rigidly rotating magnetosphere,
the stellar ARE can be detected from Earth only when the stellar magnetic dipole axis 
is almost perpendicular to the line of sight.  
This condition is realized at the rotational
phases close to the nulls of the effective magnetic field curve.
Further, the elementary ECM emission mechanism mainly amplifies
just one of the two magneto-ionic modes of the electromagnetic waves,
propagating within the magnetized plasma (each one with opposite circular polarization direction).
The growth rate of the amplified mode,
that define what mode prevails (X-mode versus O-mode), 
depends on the local magnetic field strength and plasma density 
\citep{sharma_vlahos84,melrose_etal84}.  
This explains the observed high
degree of polarization for the auroral pulses.  
The circular polarization sense of the ARE is related to the magnetic field vector
orientation of the auroral source region.  
Hence the sign of the Stokes\,{\em V} parameter bears the signature of the stellar
hemisphere where the ARE originates. 

In the case of the ARE from \roa, a helicity reversal of the L
band circular polarization sense has clearly been observed (see Fig.~\ref{fig_are_zoom}).  
The evidence of a dominant sign
of the circular polarization
at well-defined stellar orientations indicates that the beam
patterns of the auroral emission arising from the two opposite
hemispheres of \roa\ are not parallel.  
The behavior of the ARE from \roa\ is 
in accordance with the auroral beam pattern oriented at an angle lower than $90^{\circ}$ 
with respect to the local magnetic field vector. This effect could be
intrinsic to the elementary ECM amplification mechanism, or due to
the upward refraction by the dense thermal plasma trapped within
the inner-magnetosphere \citep{trigilio_etal11,lo_etal12}.

The observable effects of the auroral beam orientation { have been} extensively analyzed by \citet{leto_etal16},
who explored  the ARE features from a dipole dominated ORM magnetosphere
as a function of the beam geometry.
The model parameters that control the auroral beam pattern are:
the hollow cone thickness, angle $\Delta \theta$; the cone half-aperture, angle $\theta$;
the angular width of the beam { centred} to the plane tangent to the cavity wall. 
Assuming that the ARE from \roa\ originates from
the hotter plasma component responsible { for} the X-ray emission,
the electron energy can be used to constrain
some parameters of the beam geometry.
Following the ECM amplification mechanism, an unstable
electron population with average energy $4.1$ keV,
corresponding to the hottest plasma component used to fit the X-ray spectrum acquired 
during the strongest X-ray burst (see Table~\ref{par_xray}),
produces ARE beamed within a large conical sheet with
thickness $\Delta \theta \approx 7^{\circ}$ and half-opening angle $\theta \approx 82^{\circ}$.
The X-ray spectra model fitting have been also performed using a combination of thermal components
plus { non-thermal} X-rays photons. 
In this latter case, the temperatures of the hottest thermal components used 
to fit the three X-ray spectra were found similar during quiescent and high states. 
The ARE beam parameters corresponding to the average temperature ($kT = 2.2$ keV) are
$\Delta \theta \approx 5^{\circ}$ and $\theta \approx 84^{\circ}$.

\begin{figure}
\resizebox{\hsize}{!}{\includegraphics{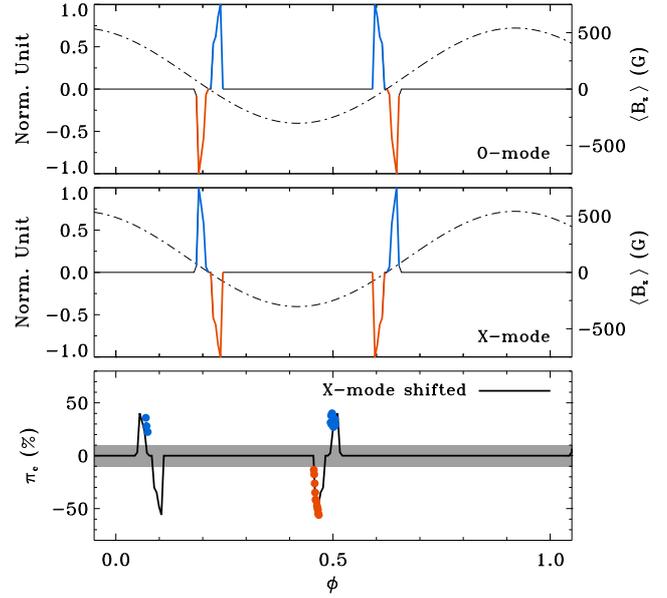}}
\caption{Synthetic ARE from \roa. Top panel: O-mode dominant magneto-ionic mode.
Middle panel: X-mode dominant mode.  The effective magnetic curve corresponding to the ORM of \roa\ is also reported (dot-dashed line).
Bottom panel: comparison between the measured circular polarization fraction (filled dots) above the detection threshold 
(grey area) with the simulated curve for the X-mode, normalized to the observed extrema.
The blue refers to the RCP polarizations sense, the red to the LCP.
To well match observations and simulations, the theoretical curve was shifted in phase,
see text for details.}
\label{sint_are}
\end{figure}

Adopting the ORM reported in Table~\ref{par_star},
we simulated the light curves of the ARE from \roa\ varying 
the extension of the auroral beam size
between 4 and 40 degrees (simulation step $2^{\circ}$),
whereas for the other model parameters
the two cases discussed above have been explored.
The adopted magnetospheric shell is constrained between $L=8$ and $L=12$ $R_{\ast}$,
in accordance with the analysis performed in Sec.~\ref{sec_magsph} and Sec.~\ref{sec_x_ray}.

The synthetic ARE produced at $\nu=2.1$ GHz was compared with the observed ones and with the 
effective magnetic curve of \roa.
The set of parameters that { best} reproduce the observed shape of the ARE from \roa\
is: beam size in the range 10--15 degrees, $\Delta \theta = 7^{\circ}$, and $\theta = 82^{\circ}$.
The identification of the above parameters suggest that the electrons radiating the ARE 
coincide with the hottest thermal component radiating X-rays, 
even if the low number of radio measurements of the ARE from \roa\
prevents a final conclusion.

The simulated auroral light curves for the cases of O (top panel) and X (middle panel) dominant mode
are pictured in Fig.~\ref{sint_are} (for the theoretical curves shown in { the figure we} used $\delta=12^{\circ}$).
The curve of the effective magnetic field, corresponding to the ORM of \roa,
is also superimposed.
The comparison between simulations and observations is shown
in the bottom panel of Fig.~\ref{sint_are}.

\begin{figure}
\resizebox{\hsize}{!}{\includegraphics{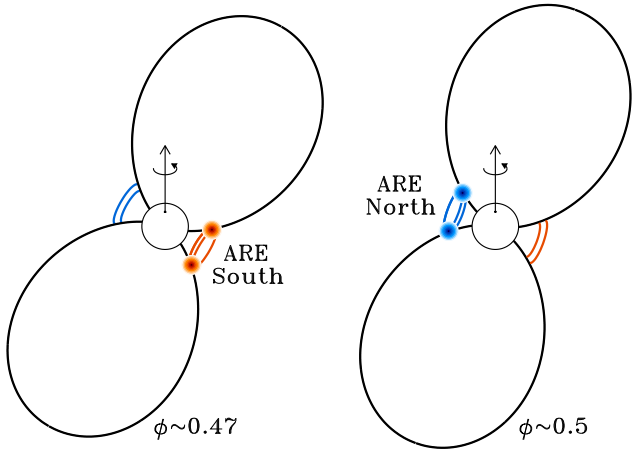}}
\caption{Meridional cross section of { \roa's} magnetosphere oriented with the magnetic axis almost perpendicular to the line of sight.
The pictured dipolar field lines are displaced in the plane of the sky.
The rings above the magnetic poles, blue for the northern and red for the southern, locate the magnetospheric region
where the second harmonic of the local gyro-frequency equals 2.5 (ring nearest to the surface) and 1.7 GHz (farthest ring).
The spots represent the highly directive auroral beam when this is almost aligned with the line of sight.
}
\label{fig_are}
\end{figure}

Comparing the sequence of the simulated auroral pulses with the measured ARE, 
we observe that the  sequence of the circularly
polarized pulses predicted by the X-mode auroral emission is { in better agreement}
with the observed one.  
Such a correlation demonstrates
that the extraordinary magneto-ionic mode is mainly amplified within
the magnetosphere of \roa.  In this case the LCP amplified emission
(Stokes\,$V<0$) mainly arises from the southern hemisphere, whereas
the RCP  emission  (Stokes\,$V>0$) originates in the northern
hemisphere (Fig.~\ref{sint_are}).  
The helicity reversal of
the ARE occurs close to $\phi \approx 0.5$, whereas the corresponding
null of the effective magnetic field  occurs close to $\phi \approx
0.6$.  Hence, 
in the layers where the ARE is produced,
the magnetic polarity reversal anticipates ($\Delta \phi \approx 0.1$) 
the corresponding null of the effective magnetic field, 
which traces the average orientation of the magnetic field vector anchored at the stellar surface.  
As already discussed at the end
of Sec.~\ref{beff_curve}, this highlights that the pure magnetic
dipole topology is only a first approximation of the true magnetic
topology of \roa.  In fact, the observed and the synthetic auroral emission
(the latter simulated assuming as a first approximation the ORM geometry retrieved in Sec.~\ref{beff_curve})
{ match well when} a phase shift is introduced; see the bottom
panel of Fig.~\ref{sint_are}.

For \roa\ the ECM
mainly amplifies the extraordinary magneto-ionic mode (bottom panel of Fig.~\ref{sint_are}), for which
the circular polarization sense corresponds with the helicity of the
electrons that move within the magnetized region.  
This is the case of the ECM operating in low-density magnetospheric regions
\citep*{sharma_vlahos84,melrose_etal84}, converse to the case of
HD\,142301 where the ECM operates in a denser region and mainly
amplifies the ordinary mode \citep{leto_etal19}.
Finally, the statistically significant detection of highly RCP
L-band emission close to the \bz\ null at $\phi \approx 0.07$ is a
further confirmation that the ARE is efficiently produced within
the magnetosphere of \roa.

The cartoon pictured in
Fig.~\ref{fig_are} shows the front view { of \roa's magnetosphere}
correctly oriented to make the stellar ARE detectable from Earth.
The spots pictured in the figure are the beams of the auroral
emission, oriented to the observer, arising, respectively, from the
northern (blue spots) and the southern (red spots) stellar hemispheres.
The non perfect perpendicular beam orientation, with respect to the local magnetic field vector,
explains the observed change of the circular polarization sense as \roa\ rotates.
The left panel of Fig.~\ref{fig_are} shows the magnetospheric orientation 
that is in accordance with the detection of the ARE from the southern hemisphere;
the right panel is instead related to ARE detection from the northern hemisphere.

As discussed above, the auroral beam orientation is related to the
elementary ECM amplification process, but further refraction effects
may be suffered by the auroral radiation { traversing} the surrounding
ionized medium.  The angle $\theta$ used to perform the simulations
discussed above was retrieved from the energy of the thermal electrons
{ responsible for} the X-ray emission from \roa.  The accordance between
simulations and observations is satisfying, { so} we conclude that
the density of the medium is low enough to not produce { significant}
further deflection or absorption effects.

\section{Discussion}
\label{discussion}
\roa\ is the sixth early-type magnetic star where a
stellar ARE has been detected  
\citep*{trigilio_etal00, das_etal18,das_etal19a,das_etal19b,leto_etal19}.  
In order for the ARE to be detected, the source must be observed when its magnetic
field axis is almost perpendicular to the line-of-sight.
But the { true magnetic field topology
could deviate from 
a simple dipole} rigidly anchored to the stellar surface.
Then, some auroral pulse might be observed shifted with respect to the phase location of the effective 
magnetic nulls. This seems to be the case in \roa\ (see bottom panel of Fig.~\ref{sint_are}).
Offset effects are commonly observed in hot magnetic stars showing coherent ARE 
\citep{kochukhov_etal14, 
leto_etal19, 
das_etal19a, 
das_etal19b}. 
Further,
the frequencies of the electromagnetic waves amplified by the ECM
are tuned at radio frequencies that are a function of the stellar magnetic field strength.  
All of these conditions make the detection of stellar ARE difficult.
Hence, the ever increasing number of early-type magnetic stars 
discovered as auroral radio sources suggest that the suitable physical conditions
able to amplify the ARE are commonly achieved within the magnetospheres of early-type stars.

The ECM emission mechanism responsible for stellar ARE in
early-type magnetospheres is likely powered by the loss-cone unstable
electron energy distribution \citep{leto_etal19}.
This unstable energy distribution can be developed if a fraction of the fast electrons, 
responsible for the incoherent non-thermal radio emission, 
deeply penetrate the magnetospheric layers impacting with the stellar surface.  
The reflected electron populations will be then deprived by those
electrons that have velocity vectors almost aligned with the
magnetic field lines giving rise to the unstable loss-cone electron
energy distribution.
The electrons' impact with the stellar surface, which is the ARE triggering event, 
could, in addition to the wind-shocks, be a significant source of X-ray emission
and also responsible for \roa's observed X-ray behavior.

Interestingly, the X-ray spectra of \roa\ are { also} compatible with 
X-ray photons with a power-law energy distribution.
The presence of non thermal X-ray photons is not  a unique prerogative of the early B magnetic stars.
The cooler star CU\,Vir 
($T_{\mathrm {eff}} \approx 13$ kK) also has an
X-ray spectrum compatible with the presence of a non-thermal component \citep{robrade_etal18}.

In the case of both \roa\ and CU\,Vir, primary (non-thermal X-ray photons) and 
secondary (coherent radio emission) observable effects 
of the electrons' impact with the stellar surface were detected.
But, at the present time, the X-rays from \roa\ 
also show evidence of a clear rotational variability and the presence of very hot thermal plasma.
If there is a significant contribution to the X-ray emission of \roa\
from the plasma process related to the fast-electrons' impact with the stellar surface,
the X-ray emitting regions will be very likely 
located in compact magnetospheric regions close to the magnetic poles.
Hence, the observed rotational modulation of the X-ray light curve might be compatible with a simple geometric effect.

\begin{figure*}
\resizebox{\hsize}{!}{\includegraphics{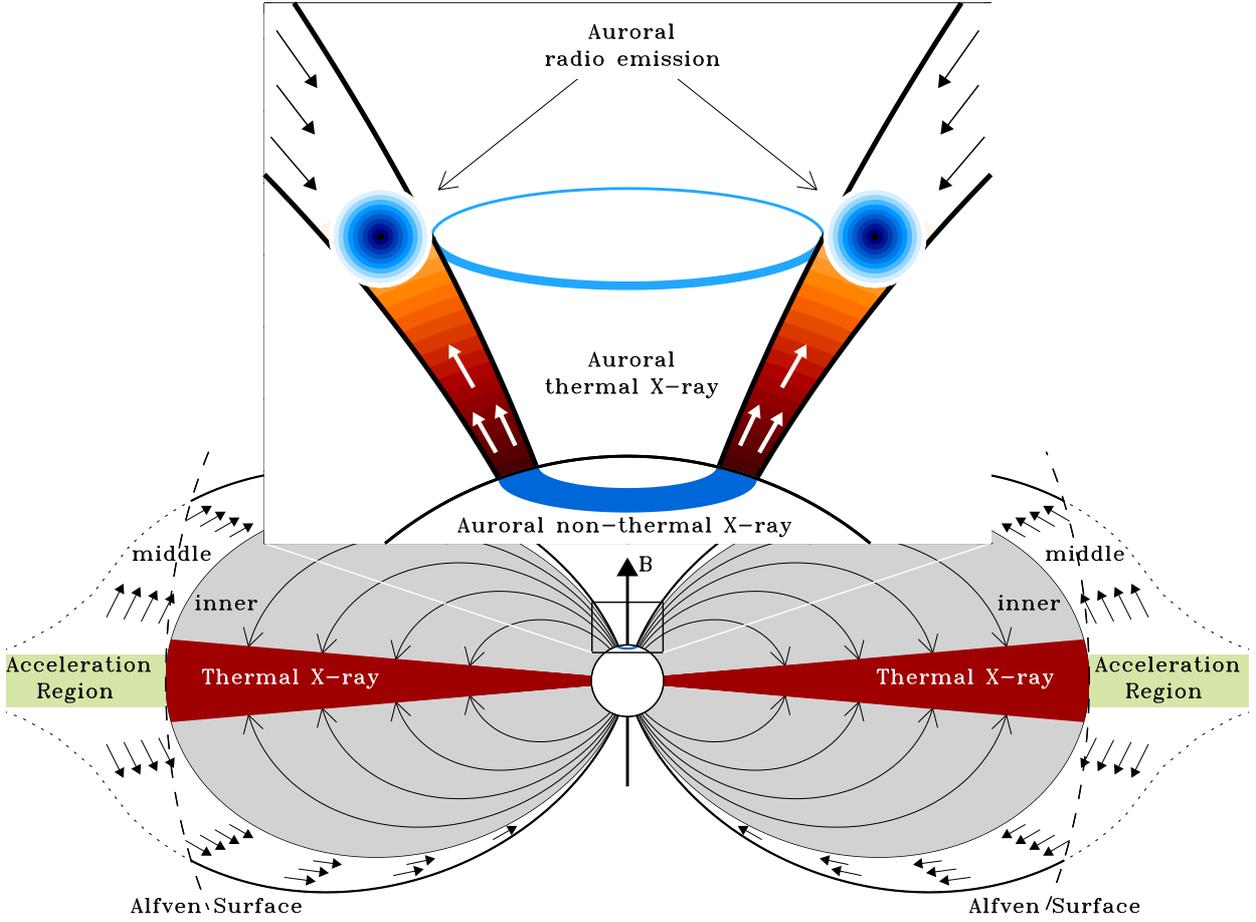}}
\caption{Sketch of the scenario proposed to consistently explain the radio and the X-ray observations of \roa.
In the magnetospheric region where the magnetic energy is strong enough the radiatively driven stellar wind 
cannot freely expand. This is the inner magnetosphere (the light grey area).
The stellar wind is channeled by the closed field lines of the magnetic dipole.
The ionized matter arising from the two magnetic hemispheres (curved arrows)
collides and shocks close to the magnetic equatorial plane,
producing thermal X-ray emission (equatorial red areas).
Far from the star, the magnetic field strength decreases. 
When the energy of the trapped matter is high enough, the trapped plasma opens the magnetic field lines forming current sheets.
These regions are sites of magnetic reconnection events that
accelerate the local plasma up to relativistic energy
(pictured using the light-green areas just outside the Alfv\'{e}n surface).
The non-thermal electrons (represented by the small solid black arrows) freely propagate within the middle-magnetosphere 
radiating at the radio regime by { the} gyro-synchrotron emission mechanism.
The non-thermal electrons that deeply penetrate the stellar magnetosphere
impact on the surface producing non-thermal X-ray photons by { the} thick target breemstrahlung emission mechanism.
The region of such non-thermal X-ray emission has a ring shape 
{ centred on the} stellar magnetic poles,
like an auroral oval.
The source of the auroral non-thermal X-ray photons (large curved blue area located on the stellar surface) 
is pictured in the zoomed view of the northern magnetosphere. 
The energy released by the bombardment of  the surface 
produces evaporation of plasma,
that fills the deep regions of the middle-magnetosphere and is { responsible for} the auroral thermal X-ray component
(areas of decreasing red intensity).
The non-thermal electrons mirrored by the converging magnetic field are deprived by those electrons 
{ that, impacting the star, originate} the auroral non-thermal X-ray photons.
This condition gives rise to { an} unstable energy distribution
in the electrons moving outwards (white solid small arrows), 
triggering the auroral coherent radio emission from rings located above the polar regions.
The anisotropic radiation beam of the ARE is almost perpendicular to the polar magnetic field vector 
and tangentially directed to the auroral ring.
The ARE beams directed to the observer are highlighted by the two blue bright spots.
}
\label{scenario}
\end{figure*}

The non-thermal plasma streams that impact the star could be a source of hot thermal plasma.
The high energy released during relativistic particle bombardment on the stellar surface
leads to plasma evaporation,
that fills the magnetic shell region close to the star of hot thermal plasma.
Such hot plasma could explain the detection of the emission line at 6.7 keV from highly ionized Fe,
that is a signature of { the} presence of plasma at temperatures beyond 60 MK.
Furthermore, a large fraction of X-rays might be created by this very hot plasma, 
producing not-negligible effects on the total X-ray continuum spectrum.

As discussed in Sec.~\ref{sec_are}, the ARE from \roa\ is in accordance with amplification
of the extraordinary magneto-ionic mode. Following the ECM theory, the X-mode is mainly amplified 
within regions where $\nu_{\mathrm p}/\nu_{\mathrm B} < 0.3$--0.35 (with $\nu_{\mathrm p}$ the local plasma frequency)
\citep*{sharma_vlahos84,melrose_etal84}. The above condition constrains the density
of the thermal electrons ($N_{\mathrm e}$) located in the region where the ARE originates.
The plasma frequency is related to $N_{\mathrm e}$ by the following relation:
$\nu_{\mathrm p} = 9 \times 10^{-6} \sqrt N_{\mathrm e}$ (GHz),
{ thus}, the electron density of the magnetospheric layer at
$\approx 1$ $R_{\ast}$ above the stellar surface (where the L-band ARE of \roa\ originates) has to be $\lessapprox 10^9$ cm$^{-3}$. 
As the radio frequency increases, the ARE arises from deeper magnetospheric layers
and the allowed plasma density is consequently higher.
The emission measure ($EM=N^2_{\mathrm e} V$, with $V$ volume of the emitting source) of this thermal plasma  
can be estimated assuming the source region constrained between 
two truncated cones, { centred} on the stellar magnetic axis. The lower and upper radii of the two truncated cones 
are given by the intersection of the  boundary magnetic field lines of the middle-magnetosphere,
with, respectively, the stellar surface and the plane perpendicular to the magnetic axis 
located above the surface where the ARE arises.
The $L$-shell parameter of the inner boundary of the middle-magnetosphere was fixed at
8 $R_{\ast}$. The outer boundary was assumed at $L = 12$ $R_{\ast}$.
This in accordance with the size of the magnetosphere of \roa\ estimated in Sec.~\ref{sec_magsph}.
Assuming conservatively a constant thermal electron density of $10^9$ cm$^{-3}$,
within the magnetospheric regions constrained between the stellar surface and the layer
where the physical conditions able to support the ARE from \roa\ are satisfied,
we estimate $\approx 10^{52}$ cm$^{-3}$
as order of magnitude of $EM$.
The above $EM$ estimation was performed imposing an electron
density compatible with the extraordinary magneto-ionic mode amplification, 
that for the electron-cyclotron maser emission mechanism
is also { expected to be} dominant  over the O-mode \citep*{melrose_etal84}.
To fit the X-ray spectrum obtained during the bright states 
we need $EM \approx 10^{53}$ cm$^{-3}$ for the hottest thermal
plasma component (see Table~\ref{par_xray}).  To produce a stronger
X-ray emission from the auroral regions, a hot plasma with density
higher than $10^9$ cm$^{-3}$ is required.  But, such high density
plasma inhibits the X-mode propagation, so the ARE from \roa\
will { expected to be} strongly suppressed.

The auroral X-ray photons might have a dual origin:
non-thermal emission from the auroral oval; and
thermal emission from the very hot plasma just above the surface. 
In this case ARE and X-ray emissions are expected to anti correlate.
When the plasma conditions are suitable for ARE amplification and propagation,
the consequent auroral X-ray emission is expected to be quite low.   
On the contrary, higher plasma density within the auroral source regions 
is compatible with the observed X-ray bright states,
but the coherent auroral radio emission might be strongly inhibited.
We showed that 
the visibility of the magnetospheric regions where the incoherent radio emission is maximum,
nearly corresponding to a pole-on view of the magnetic poles,
may be a favorable geometry for detecting the thermal auroral X-rays
(see bottom panel of Fig.~\ref{fig_model_x}).
But at the present, we cannot { be sure} if only the stellar rotation is
able to produce observable variability effects on the thermal auroral
X-ray emission from \roa.

It is worth { noting} that
the power of the radiatively driven wind of \roa\ (higher than  $10^{32}$ erg s$^{-1}$, Sec.~\ref{sec_magsph}) 
is higher than the cumulative power radiated at the radio ($\approx 1.8 \times 10^{29}$ erg s$^{-1}$, Sec.~\ref{radio_incoe})
and X-ray ($\lessapprox 1.9 \times 10^{31}$ erg s$^{-1}$, Sec.~\ref{sec_x_ray}), hence,
the wind energy of \roa\ is high enough to power both radio and X-ray emissions.
Extending the MCWS scenario,
the stellar wind might be the engine of plasma processes responsible for 
{ \roa's emission}, from the radio to the X-ray, including those of auroral origin, Fig.~\ref{scenario}.

On the basis of the common observing features that characterize different hot magnetized stars,
the physical processes acting within the magnetosphere of \roa\ are very likely 
acting within the magnetospheres of all the stars belonging to this class.
Obviously, the observing features induced by the plasma processes { described above} will be seen 
in different ways from different objects.
For example, stellar geometry and rotation inclination has a crucial role for ARE detectability.

However, open questions remain.
The \XMM\ measurements covered a time range 140 ks long ($\approx 1.6$ days),
that corresponds to $\approx 2$ full rotations of \roa\ ($P_{\mathrm{rot}} \approx 0.75$ days).
During the 2016 X-ray measurements \roa\ showed each magnetic pole two times.
From a pure geometrical point of view, the detection of four X-ray pulses was expected,
while only two were detected,
meaning that not all polar passages produce X-ray bright states.

As discussed above, the triggering source of the auroral radio and X-ray phenomena are the non-thermal electrons
responsible for the incoherent gyro-synchrotron radio emission.
The stability of the observed flux levels ensures a constant density of these relativistic electrons 
within the middle-magnetosphere of \roa, at least over the 5 days covered by the ATCA measurements.
But we reasonably expect the time scale over which the radio emission of \roa\ is stable is much longer than five days.
In fact, multi epoch observations showed that the radio emission of other early-type magnetic stars
(i.e. $\sigma$\,Ori\,E and HD\,142301) was stable over a time scale of years \citep{leto_etal12,leto_etal19}.
This implies that magnetic reconnection, which is the acceleration mechanism able to create these fast electrons,
likely occurs as a steady mechanism within the magnetospheres of the early-type magnetic stars.
This contrasts with the case of flares from the Sun and other cool stars,  
where magnetic reconnection appears more stochastic.

The plasma process occurring within the magnetospheres of the
early-type magnetic stars are triggered by their stellar winds.  
The mass loss of such kind of main sequence stars is uniform and steady,
hence their magnetospheric physical conditions are quite stable.
Nevertheless, there is some evidence that the ARE behavior can be variable.
For CU\,Vir, the pulses do not exactly repeat, although they do
show a regular phase arrival time \citep{lo_etal12,pyper_etal13}.
Consequently, the ARE of hot stars is not necessarily stable. The
fast electron population causing the stable incoherent non-thermal
radio emission from \roa\ also induces auroral phenomena, but these
have not been observed to be as stable.  The ORM of \roa\ is
compatible with two auroral pulses per period, but we clearly
detected only one pulse. In accordance with the ORM geometry of
\roa, at the phases where the second auroral pulse is expected
there is only a tentative detection, this supports the ORM geometry
of \roa, but meantime evidenced that the ARE of \roa\ has a variable flux.  
In accordance with the proposed common origin for the radio
and X-ray auroral phenomena, the non-thermal particle bombardment
of the stellar surface produces evaporation of plasma and a high
level of X-ray emission, 
but meantime the presence of such hot thermal plasma increases
the density of thermal matter within the deep layers of the middle magnetosphere, 
that opposes the deep penetration of the continuous flux of precipitating 
non-thermal particles close to the polar regions of the stellar surface.
The inability of the reflected { electron} population to develop 
the loss-cone unstable energy distribution shuts down the ARE.
When the hot plasma has rarefied, the duty cycle for the radio and X-ray auroral
phenomena restarts.
The above discussion is merely a speculation
based only { on} the empirical fact that the auroral radio and X-ray
emissions of \roa\ were not always detected (or just tentatively
detected) when the suitable geometric conditions are verified.

\section{Conclusions and outlook}
\label{conclusion}
In this paper we report multi-wavelength  ATCA measurements of the
early-type magnetic star \roa.  The ATCA radio observations of \roa\
reveal that two distinct radio emission processes are active.
The measurements in the frequency range 5.5--21.2 GHz are compatible with
incoherent gyro-synchrotron { radiation} from a stable co-rotating magnetosphere.
The 2.1~GHz ATCA measurements are characterized by two highly polarized
pulses.  The  low-frequency behavior clearly indicates that the
coherent stellar Auroral Radio Emission was detected.

High-resolution optical spectra of \roa, obtained by several instruments, have
been collected. 
Further, new magnetic field measurements of \roa\ { were} also reported.
The rotational variability of the radio and optical observables 
allowed { determination of} the rotation period of of \roa.
The rotation period was used to obtain
the effective magnetic curve of \roa, using already published
\citep{pillitteri_etal18} with new magnetic field measurements. The
geometry of the oblique rotator model that  represents the magnetic
field topology of \roa\ was inferred.
Using this ORM geometry, the measured incoherent and coherent 
radio emissions of \roa\ were { modelled allowing us} to constrain some physical 
conditions of the stellar magnetosphere 
(i.e. the size of the magnetosphere; the wind mass loss rate).

\roa\ is characterized by strong and variable X-ray emission
\citep{pillitteri_etal17}.  
The X-ray data 
reveal the presence of a very hot thermal plasma. 
{ Further, the} X-ray spectra 
are compatible with the presence of non-thermal X-ray photons.
These X-ray features are not compatible with the standard 
MCWS model.
The behavior of the X-ray emission of \roa\
suggests an auroral origin,
where non-thermal X-rays and very hot thermal plasma 
might be generated from relativistic electrons impacting the surface.

\roa\ is not the first early-type magnetic star showing a possible non-thermal
X-ray spectral component.  There are three other hot magnetic
stars showing X-ray spectra compatible with a non-thermal component
\citep{leto_etal17a,leto_etal18,robrade_etal18}.  
However \roa\ 
is the first to display a clear rotational modulation of X-rays in
accordance with visibility of the magnetic poles. 
Furthermore, the ratio
between the X-ray uminosity and the radio spectral luminosity of \roa\ is about an order
of magnitude higher compared to ratios measured for the other early-type
magnetic stars where there are hints
of auroral X-ray emission \citep{robrade_etal18}. 
The auroral X-ray components of \roa\ are then significant sources of X-rays
of comparable strength with the X-rays of wind-shock origin. 

We conclude that
the stellar wind is the engine that powers the plasma processes responsible for the observed 
radio and X-ray behavior of \roa. 
The scenario provided by the MCWS model that includes the auroral phenomena is outlined in Fig.~\ref{scenario}.  
The acceleration process
occurring in the farther regions of the stellar magnetosphere,
that produces the non-thermal plasma responsible for the
incoherent gyro-synchrotron emission,
are also able
to drive the auroral phenomena observed at the radio and X-ray wavelengths.
Similar physics is likely operating within the magnetospheres of all
stars belonging to this class.  
{ The insights} obtained from the study of \roa\ { might therefore have} general applications.

Finally, we observed that the radio and X-ray auroral features of \roa\
were not always detected as expected.
To investigate further the physical mechanisms 
responsible for the auroral phenomena which
induce the observed radio/X-ray behavior of \roa,
new and coordinated radio and X-ray observations, 
covering more than one rotation period, will be crucial.

The challenge with detecting auroral X-ray emission from magnetic massive stars is that both 
the non-thermal electron streams colliding with the stellar surface 
and the standard embedded wind shocks, occurring in the outflowing wind from the hemispheres 
of opposite magnetic polarity, produce thermal X-rays. 
Consequently, the non-thermal X-rays of auroral origin 
compete with these thermal components for identification via spectral modeling.
In general, X-ray measurements covering
a wide energy range would be useful to disentangle the { emission contributions} from
all these different plasma components.

\section*{Acknowledgments}
We sincerely thank Dr. Matthew Shultz, who reviewed this paper
giving us very useful and constructive criticisms, comments, and suggestions,
that helped to significantly improve the paper.
This work has extensively used the NASA's Astrophysics Data System, and the 
SIMBAD database, operated at CDS, Strasbourg, France. 
This work is based 
on observations collected at the European Organisation 
for Astronomical Research in the Southern Hemisphere under ESO programme(s) 101.D-0131(A), and,
on observations made with the Italian Telescopio Nazionale Galileo (TNG) 
operated on the island of La Palma by the Fundaci\'{o}n Galileo Galilei of the INAF 
(Istituto Nazionale di Astrofisica) at the Spanish Observatorio 
del Roque de los Muchachos of the Instituto de Astrofisica de Canarias.
LMO acknowledges support from
the DLR under grant FKZ\,50\,OR\,1809 and partial support by the Russian
Government Program of Competitive Growth of Kazan Federal University.
JK was supported by grant GA\,\v{C}R 18-05665S.



\appendix
\section{List of radio and EW measurements}


\begin{table} 
\caption{{ First 5 rows of the radio data for each observing band. The complete list of the ATCA measurements is available online.}} 
\label{radio_data} 
\begin{tabular}{@{}ccc @{~~}c@{~~} ccc@{}} 
\hline 
HJD         &$S_{\mathrm I}$                 &$S_{\mathrm V}$          &    &HJD         &$S_{\mathrm I}$                 &$S_{\mathrm V}$            \\
2450000+    &mJy                             &mJy                                     &    &2450000+    &mJy                             &mJy                        \\
\hline 

\cline{1-3}  
                        \cline{5-7} 
\multicolumn{3}{c}{L band {\s ($\nu=2.1$ GHz)}}                             &    &\multicolumn{3}{c}{U band {\s ($\nu=16.7$ GHz)}}                                      \\                
\cline{1-3}  
                        \cline{5-7}  

8555.0683      &4.8 \s{($\pm0.3$)}    & $ 0.1$  \s{($\pm 0.3$)}   &    & 8555.1015     &6.7 \s{($\pm0.4$)}      & 1.0  \s{($\pm0.4$)} ~~~ \\
8555.0691      &5.3 \s{($\pm0.3$)}    & $ 0.4$  \s{($\pm 0.3$)}   &    & 8555.1022     &6.4 \s{($\pm0.4$)}      & 1.2  \s{($\pm0.4$)} ~~~ \\
8555.0699      &5.6 \s{($\pm0.3$)}    & $ 0.0$  \s{($\pm 0.3$)}   &    & 8555.1030     &6.5 \s{($\pm0.4$)}      & 1.4  \s{($\pm0.4$)} ~~~ \\
8555.0706      &4.3 \s{($\pm0.3$)}    & $ 0.0$  \s{($\pm 0.3$)}   &    & 8555.1038     &5.8 \s{($\pm0.4$)}      & 1.3  \s{($\pm0.4$)} ~~~ \\
8555.0714      &4.5 \s{($\pm0.3$)}    &$-0.1$  \s{($\pm 0.3$)}~~~   &    & 8555.1045     &6.3 \s{($\pm0.4$)}      & 1.2  \s{($\pm0.4$)} ~~~ \\

\cline{1-3}  
                        \cline{5-7} 
\multicolumn{3}{c}{C band {\s ($\nu=5.5$ GHz)}}                             &    &\multicolumn{3}{c}{K band {\s ($\nu=21.2$ GHz)}}                                      \\
\cline{1-3}  
                        \cline{5-7} 
8555.0827      &7.1 \s{($\pm0.3$)}     & 0.5  \s{($\pm0.3$)}   &    &  8555.1015     & 6.3  \s{($\pm0.6$)}      & 2.2 \s{($\pm0.6$)} ~~~ \\
8555.0835      &7.0 \s{($\pm0.3$)}     & 0.9  \s{($\pm0.3$)}   &    &  8555.1022     & 5.0  \s{($\pm0.6$)}      & 2.0 \s{($\pm0.6$)} ~~~ \\
8555.0843      &6.4 \s{($\pm0.3$)}     & 0.7  \s{($\pm0.3$)}   &    &  8555.1030     & 5.2  \s{($\pm0.6$)}      & 1.1 \s{($\pm0.6$)} ~~~ \\
8555.0851      &6.4 \s{($\pm0.3$)}     & 1.2  \s{($\pm0.3$)}   &    &  8555.1038     & 5.5  \s{($\pm0.6$)}      & 0.7 \s{($\pm0.6$)} ~~~ \\
8555.0858      &6.5 \s{($\pm0.3$)}     & 1.1  \s{($\pm0.3$)}   &    &  8555.1045     & 5.2  \s{($\pm0.6$)}      & 1.2 \s{($\pm0.6$)} ~~~ \\

\cline{1-3}   
\multicolumn{3}{c}{X band {\s ($\nu=9$ GHz)}}                               &    &            &                               &                                   \\
\cline{1-3}   
 8555.0827   &8.2 \s{($\pm0.3$)}    &1.1 \s{($\pm0.3$)}      &    &            &                               &                                   \\
 8555.0835   &8.2 \s{($\pm0.3$)}    &1.4 \s{($\pm0.3$)}      &    &            &                               &                                   \\
 8555.0843   &8.1 \s{($\pm0.3$)}    &1.3 \s{($\pm0.3$)}      &    &            &                               &                                   \\
 8555.0851   &7.7 \s{($\pm0.3$)}    &1.4 \s{($\pm0.3$)}      &    &            &                               &                                   \\
 8555.0858   &7.9 \s{($\pm0.3$)}    &1.4 \s{($\pm0.3$)}      &    &            &                               &                                   \\

\hline 

\end{tabular}     
\end{table}



\begin{table} 
\caption{EW measurements of the He\,{\sc i} line at  $\lambda=5015$~\AA.} 
\label{ew_data} 
\begin{tabular}{@{}cc @{~~}c@{~~} cc@{}} 
\hline 
HJD        &EW                                           &   &HJD        &EW                                         \\
2450000+   &\AA\                                         &   &2450000+   &\AA\                                       \\
\hline 
\cline{1-2}                                              
                                                             \cline{4-5}                                                                                  
\multicolumn{2}{l}{HiRes: {\s range 4400--6800 \AA, R=57\,000}}     &   &\multicolumn{2}{l}{NARVAL: {\s range 3700--10500 \AA, R=80\,000}}       \\            
                                                             \cline{4-5}                                                              
\cline{1-2}
~672.784     &312 \s{($\pm20$)}                          &   & 6849.372   &312 \s{($\pm10$)}                             \\     
\cline{1-2}                                                                        
\multicolumn{2}{l}{UVES: {\s range 4800--6800 \AA, R=74\,000}}          &   & 6849.387   &323 \s{($\pm15$)}                             \\
\cline{1-2}                                                                     
2132.464    &382 \s{($\pm20$)}                           &   & 6849.403   &318 \s{($\pm20$)}                             \\ 
2132.660    &420 \s{($\pm30$)}                           &   & 6849.419   &319 \s{($\pm15$)}                             \\
                                                             \cline{4-5}                                                       
2133.464    &408 \s{($\pm25$)}                           &   &\multicolumn{2}{c}{CAOS: {\s range 3750--11000 \AA, R=45\,000}}      \\
                                                             \cline{4-5}                          
7881.646    &364 \s{($\pm10$)}                           &   & 8262.493   &336 \s{($\pm35$)}                            \\                                                                                                     
7920.489   &365 \s{($\pm10$)}                           &   & 8268.476   &341 \s{($\pm40$)}                           \\
\cline{1-2}                                                                                                            
\multicolumn{2}{l}{ESPaDOnS: {\s range 3700--10500 \AA, R=80\,000}}           &   & 8269.462   &394 \s{($\pm40$)}                         \\
\cline{1-2}                                                                
3511.944        &325 \s{($\pm10$)}                     &   & 8270.507   &391 \s{($\pm35$)}                         \\
\cline{1-2}                                                              
\multicolumn{2}{l}{HARPS: {\s range 3800--6900 \AA, R=115\,000}}         &   & 8290.426   &387 \s{($\pm40$)}                         \\
\cline{1-2}                                             
4189.804      &345 \s{($\pm20$)}                         &   & 8291.370   &388 \s{($\pm50$)}                          \\
4189.828       &368 \s{($\pm20$)}                        &   & 8299.377   &381 \s{($\pm50$)}                            \\
\cline{1-2}                                                                                     
\multicolumn{2}{l}{HARPS-N: {\s range 3800--6900 \AA, R=115\,000}}       &   & 8306.369   &393 \s{($\pm60$)}                            \\
\cline{1-2}                                                 
8295.443      &310 \s{($\pm30$)}                        &   &            &                                               \\
\hline                                                                                                             

\end{tabular}     
\end{table}


\end{document}